\documentclass[letterpaper,american,reprint, aps, prc, latin1, superscriptaddress]{revtex4-2}
\usepackage{mathptmx}
\usepackage[T1]{fontenc}
\usepackage[latin9]{inputenc}
\usepackage{color}
\usepackage{babel}
\usepackage{array}
\usepackage{units}
\usepackage{multirow}
\usepackage{amsmath}
\usepackage{amssymb}
\usepackage{graphicx}
\usepackage{esint}
\usepackage[unicode=true,pdfusetitle,
 bookmarks=true,bookmarksnumbered=false,bookmarksopen=false,
 breaklinks=false,pdfborder={0 0 0},pdfborderstyle={},backref=false,colorlinks=true]
 {hyperref}

\makeatletter

\pdfpageheight\paperheight
\pdfpagewidth\paperwidth

\providecommand{\tabularnewline}{\\}

\newcommand{\nuc}[2]{$^{#2}$#1}
\newcommand{\amev}[1]{{#1} MeV/nucleon}
\newcommand{\collision}[5]{\nuc{#1}{#2}+\nuc{#3}{#4} collisions at \amev{#5}}

\newcommand{\col}[4]{\nuc{#1}{#2}+\nuc{#3}{#4}}

\newcommand{\xesn}{\col{Xe}{129}{Sn}{nat}}
\newcommand{\etlcp}{$E_{\mathrm{t12}}$}
\newcommand{\mtot}{$N_{\mathrm{C}}$}
\newcommand{\meanetlcp}{$\overline{E_{\mathrm{t12}}}$}
\newcommand{\meanmtot}{$\overline{N_{\mathrm{C}}}$}
\newcommand{\arni}{\col{Ar}{36}{Ni}{58}}
\newcommand{\arkcl}{\col{Ar}{36}{KCl}{}}
\newcommand{\nini}{\col{Ni}{58}{Ni}{58}}
\newcommand{\niau}{\col{Ni}{58}{Au}{197}}
\newcommand{\xesngsi}{\col{Xe}{129}{Sn}{124}}
\newcommand{\anyxesn}{\col{Xe}{129}{Sn}{x}}
\newcommand{\auau}{\col{Au}{197}{Au}{197}}

\newcommand{\ganil}{Grand Acc\'{e}l\'{e}rateur National d'Ions Lourds (GANIL), UPR 3266, CEA-DRF/CNRS-IN2P3, Bvd. Henri Becquerel, 14076 Caen Cedex, France}
\newcommand{\lpccaen}{Normandie Univ, ENSICAEN, UNICAEN, CNRS/IN2P3, LPC Caen, F-14000 Caen, France}
\newcommand{\subatech}{SUBATECH UMR 6457, IMT Atlantique, Universit\'{e} de Nantes, CNRS-IN2P3, F-44300 Nantes, France}
\newcommand{\ijclab}{Universit\'{e} Paris-Saclay, CNRS/IN2P3, IJCLab, 91405 Orsay, France}
\newcommand{\cherbourg}{Ecole des Applications Militaires de l'Energie Atomique, BP 19 50115, Cherbourg Arm\'{e}es, France}
\newcommand{\napoli}{Dipartimento di Fisica 'E. Pancini' and Sezione INFN, Universit\`{a} di Napoli 'Federico II', I-80126 Napoli, Italy}

\makeatother

\begin{document}
\title{Model independent reconstruction of impact parameter distributions
for intermediate energy heavy ion collisions}
\author{\selectlanguage{american}%
J.D.~Frankland}
\thanks{\selectlanguage{american}%
Corresponding author}
\email{john.frankland@ganil.fr}

\affiliation{\selectlanguage{american}%
\ganil}
\author{\selectlanguage{american}%
D.~Gruyer}
\affiliation{\selectlanguage{american}%
\lpccaen}
\author{\selectlanguage{american}%
E.~Bonnet}
\affiliation{\selectlanguage{american}%
\subatech}
\author{\selectlanguage{american}%
B.~Borderie}
\affiliation{\selectlanguage{american}%
\ijclab}
\author{\selectlanguage{american}%
R.~Bougault}
\affiliation{\selectlanguage{american}%
\lpccaen}
\author{\selectlanguage{american}%
A.~Chbihi}
\affiliation{\selectlanguage{american}%
\ganil}
\author{\selectlanguage{american}%
J.E.~Ducret}
\affiliation{\selectlanguage{american}%
\ganil}
\author{\selectlanguage{american}%
D.~Durand}
\affiliation{\selectlanguage{american}%
\lpccaen}
\author{\selectlanguage{american}%
Q.~Fable}
\affiliation{\selectlanguage{american}%
\lpccaen}
\author{\selectlanguage{american}%
M.~Henri}
\affiliation{\selectlanguage{american}%
\ganil}
\author{\selectlanguage{american}%
J.~Lemari\'{e}}
\affiliation{\selectlanguage{american}%
\ganil}
\author{\selectlanguage{american}%
N.~{Le Neindre}}
\affiliation{\selectlanguage{american}%
\lpccaen}
\author{\selectlanguage{american}%
I.~Lombardo}
\affiliation{\selectlanguage{american}%
INFN Sezione di Catania, via Santa Sofia 64, I-95123 Catania, Italy}
\author{\selectlanguage{american}%
O.~Lopez}
\affiliation{\selectlanguage{american}%
\lpccaen}
\author{\selectlanguage{american}%
L.~Manduci}
\affiliation{\selectlanguage{american}%
\lpccaen}
\affiliation{\selectlanguage{american}%
\cherbourg}
\author{\selectlanguage{american}%
M.~P\^{a}rlog}
\affiliation{\selectlanguage{american}%
\lpccaen}
\affiliation{\selectlanguage{american}%
Horia Hulubei National Institute for R\&D in Physics and Nuclear Engineering
(IFIN-HH), P.O.BOX MG-6, RO-76900 Bucharest-M\`{a}gurele, Romania}
\author{\selectlanguage{american}%
J.~Quicray}
\affiliation{\selectlanguage{american}%
\lpccaen}
\author{\selectlanguage{american}%
G.~Verde}
\affiliation{\selectlanguage{american}%
INFN Sezione di Catania, via Santa Sofia 64, I-95123 Catania, Italy}
\affiliation{\selectlanguage{american}%
Laboratoire des 2 Infinis - Toulouse (L2IT-IN2P3), Universit\'{e}
de Toulouse, CNRS, UPS, F-31062 Toulouse Cedex 9, France}
\author{\selectlanguage{american}%
E.~Vient}
\affiliation{\selectlanguage{american}%
\lpccaen}
\author{\selectlanguage{american}%
M.~Vigilante}
\affiliation{\selectlanguage{american}%
\napoli}
\collaboration{\selectlanguage{american}%
INDRA Collaboration}
\noaffiliation
\date{\today}
\begin{abstract}
We present a model-independent method to reconstruct the impact parameter
distributions of experimental data for intermediate energy heavy ion
collisions, adapted from a recently proposed approach for ultra-relativistic
heavy ion collisions. The method takes into account the fluctuations
which are inherent to the relationship between any experimental observable
and the impact parameter in this energy range. We apply the method
to the very large dataset on heavy ion collisions in the energy range
\amev{20\textendash 100} obtained with the INDRA multidetector since
1993, for two observables which are the most commonly used for the
estimation of impact parameters in this energy range. The mean impact
parameters deduced with this new method for ``central'' collisions
selected using typical observable cuts are shown to be significantly
larger than those found when fluctuations are neglected, and as expected
the difference increases as bombarding energy decreases. In addition,
we will show that this new approach may provide previously inaccessible
experimental constraints for transport models, such as an estimation
of the extrapolated mean value of experimental observables for $b=0$
collisions. The ability to give more realistic, model-independent,
estimations of the impact parameters associated to different experimental
datasets should improve the pertinence of comparisons with transport
model calculations which are essential to better constrain the equation
of state of nuclear matter. 
\end{abstract}
\maketitle

\section{Introduction}

The equation of state (EoS) of bulk nuclear matter in a wide range
of densities, temperatures, and proton-neutron asymmetries is of major
importance not only for nuclear physics but also  astrophysics since
the EoS plays a fundamental role in the understanding of core-collapse
supernovae (CCSN), proto-neutron star cooling \citep{Lattimer2000Nuclear},
and neutron star mergers as recently observed through gravitational
wave data \citep{Abbott2017}. In the laboratory, precise constraints
on the finite temperature EoS away from saturation density can be
obtained from heavy-ion collisions (HIC), which can be used to explore
a wide range of density, energy and asymmetry conditions, depending
on the bombarding energy, neutron and proton numbers ($N$, $Z$)
of projectile and target, and impact parameter, $b$. In particular,
the freeze-out stage of central HIC in the energy range \amev{20\textendash 100}
produces transient finite systems with similar temperatures and densities
as CCSN matter, albeit with smaller asymmetries \citep{Horowitz2014Way,Pais2019}.

The EoS cannot be measured directly but has to be inferred by comparing
the outcome of carefully selected collisions with model predictions.
For example, transport models \citep{Gregoire1987Semiclassical,Bertsch1988Guide,Aichelin1991Quantum,Ono2013,Ayik1992BoltzmannLangevin,Bonasera1994,Ono2004a,Buss2012Transporttheoretical,Napolitani2013Bifurcations,Lin2019a}
can be used to predict the dynamics of collisions at different impact
parameters using different forms of the EoS, by varying the parameters
of the employed effective interaction. For such a comparison to be
meaningful, the calculations should ideally be run over the same impact
parameter distribution $P(b|\mathbb{S})$ as that of the experimental
event sample $\mathbb{S}$, or at least use a representative value
such as the mean of this distribution, $\langle b\rangle_{\mathbb{S}}$.
This is all the more important as currently transport model calculations
are not only dependent on the EoS parameters, but also on many other
ingredients (related to uncertainties both of physics and of numerical
implementation) which are as yet not fully under control \citep{Xu2016Understanding,Zhang2018Comparison},
therefore reducing trivial bias due to mismatching of experimental
and simulated impact parameters is essential to make progress.

Of course, the impact parameter for each collision cannot be measured
but only inferred from the final state observables of each event.
An essential feature of HIC in the \amev{20\textendash 100} energy
range are fluctuations. The fluctuations of any observable $X$ from
one collision to the next can be of the same order of magnitude as
its mean value, $\langle X\rangle$. Due to a combination of diminishing
cross-section and increasing importance of fluctuations as $b\rightarrow0$,
higher and higher cuts in any observable, even one strongly correlated
with $b$, will always lead to mixing between collisions over a broad
range of impact paramter \citep{Peter1990,Phair1992Impactparameter}.
Impact parameter mixing in experimental data samples of ``central''
collisions is a general feature which can seriously bias comparisons
with transport models if neglected \citep{Zhang2011Unified}. For
example, the authors of \citep{Li2018} studied how mixing can affect
studies of isospin sensitive variables using the ImQMD transport model.
They show that in reality the mean impact parameter selected by higher
and higher cuts on the total charged particle multiplicity ``saturates''
at a finite value $b>0$, and the difference between estimated and
true centrality worsens as bombarding energies decrease below \amev{~70}
(see also for example \citep{Nebauer1999Multifragmentation}). Given
the current uncertainties in existing transport models, it would be
an important advance to be able to quantitatively characterize the
centrality of selected event samples in a model-independent way.

There are many works in the literature dealing with the characterization
of experimental centrality, from the simplest and most widely used
geometric approximation of \citep{Cavata1990Determination} to an
AI approach using neural networks trained by simulated data \citep{Haddad1997}.
Recently, a new method for reconstructing experimental impact parameter
distributions was proposed for ultra-relativistic collisions in \citep{Das2018Relating,Rogly2018Reconstructing}.
It is model independent and explicitly takes into account the fluctuations
in the relationship between $X$ and $b$, which is adjusted in order
to reproduce the experimentally measured inclusive $P(X)$ distribution.
Despite the orders of magnitude differences in beam energies and physics,
the inclusive $P(X)$ distributions for different observables used
to gauge the centrality of \emph{e.g.} Au+Au collisions at center
of mass (c.m.) energies $\sqrt{s}$=130 GeV or\emph{ }Pb+Pb collisions
at $\sqrt{s}=$2.76 TeV (see Fig. 1 of \citep{Das2018Relating}) have
generic properties very similar to those seen for e.g. total multiplicity
or total transverse energies in HIC collisions at intermediate energies:
the highest cross-section for the smallest values of $X$, decreasing
to a wide plateau and finally a near-exponential fall-off for the
largest $X$ values. Indeed such a generic distribution is expected
for any observable whose mean value decreases monotonically with $b$
when weighted with a geometric impact parameter distribution, $P(b)\sim b$.

In this paper, our aim is not to criticize or improve existing methods
of \emph{selecting} experimental samples of central collisions. Rather,
we propose a new method to \emph{characterize} any selected set of
experimental data in terms of the corresponding impact parameter distribution.
The method is model independent and uses experimental data as its
sole input. It is based on the approach of \citep{Das2018Relating,Rogly2018Reconstructing}
which we will first present along with the adaptations we have made
for its use in the intermediate energy range. Before applying it to
our experimental data, we will first of all validate the method with
a full simulation of a typical near-Fermi energy heavy-ion reaction
using the microscopic transport model AMD \citep{Ono2019}, the statistical
decay code GEMINI++ \citep{Charity2010Systematic,Mancusi2010Unified}
and a software filter reproducing the characteristics of the INDRA
array \citep{KaliVeda,jdfrankland:Pouthas1995INDRA}. Then we will
apply it to the very large INDRA dataset on heavy ion collisions in
the energy range \amev{20\textendash 100}. We will use it to provide
a systematic estimate of the true centrality of the collisions selected
by a high-transverse energy cut, as often used in previous analyses
\citep{Frankland2005}. In conclusion, by providing more reliable,
model-independent estimates of impact parameters associated to different
experimental data, this method will improve the quality of data-model
comparisons which are essential to better constrain the equation of
state of nuclear matter.

\section{Method\label{sec:Method}}

\subsection{General approach\label{subsec:General-approach}}

Consider an observable $X$ whose functional dependence on the impact
parameter can be written in terms of a conditional probability distribution
$P(X|b)$, which encodes both the $b$-dependence of the mean value,
$\overline{X}(b)$, and the fluctuations of $X$ about this mean value.
The inclusive distribution of $X$ resulting from all measured collisions,
having an unknown impact parameter distribution $P(b)$, is given
by
\begin{equation}
P(X)=\int_{0}^{\infty}P(b)\,P(X|b)\,\mathrm{d}b\label{eq:integrate-p(x|b)-over-b}
\end{equation}
Introducing the quantity \emph{centrality,} $c_{b}$, defined as the
cumulative distribution function of $P(b)$,
\begin{equation}
c_{b}\equiv\int_{0}^{b}P(b')\,\mathrm{d}b'\label{eq:b-centrality}
\end{equation}
Eq. \ref{eq:integrate-p(x|b)-over-b} can be rewritten as
\begin{equation}
P(X)=\int_{0}^{1}P(c_{b})\,P(X|c_{b})\,\mathrm{d}c_{b}=\int_{0}^{1}P(X|c_{b})\,\mathrm{d}c_{b}\label{eq:int-over-cb-p(x)}
\end{equation}
as by definition $P(c_{b})=1,\:\forall c_{b}$, and the dependency
on the unknown inclusive impact parameter distribution disappears.
Then Eq. \ref{eq:int-over-cb-p(x)} can in principle be used to determine
$P(X|c_{b})$ by fitting the experimentally measured inclusive distribution
of $X$, $P(X)$. In order to make the problem tractable, the authors
of \citep{Das2018Relating,Rogly2018Reconstructing} proposed to write
$P(X|c_{b})$ as

\begin{equation}
P(X|c_{b})=f\left[\overline{X}(c_{b}),\theta(c_{b})\right]\label{eq:fluctuation-kernel}
\end{equation}
where $f$ is a suitable p.d.f. with mean value $\overline{X}(c_{b})$
and reduced variance determined by $\theta(c_{b})=\mathrm{var}(X)/\overline{X}(c_{b})$.
In principle, both the mean value and the variance of the observable
can depend on centrality, i.e. the impact parameter.

Once $P(X|c_{b})$ is determined by fitting the experimental $P(X)$
distribution, it can be used to estimate the centrality distribution
for any given selection of data. For example, for a cut such as $x_{1}\leq X\leq x_{2}$
the corresponding centrality distribution is given by
\[
P(c_{b}|x_{1}\leq X\leq x_{2})=\frac{\int_{x_{1}}^{x_{2}}P(c_{b},X)\:\mathrm{d}X}{\int_{x_{1}}^{x_{2}}P(X)\:\mathrm{d}X}=\frac{\int_{x_{1}}^{x_{2}}P(c_{b}|X)P(X)\:\mathrm{d}X}{\int_{x_{1}}^{x_{2}}P(X)\:\mathrm{d}X}
\]
where, in the first integral, $P(c_{b},X)$ is the joint probability
distribution, $P(c_{b},X)=P(c_{b}|X)P(X)=P(X|c_{b})P(c_{b})$. The
second equality being nothing but Bayes' theorem, we can use it to
rewrite the integrand in the numerator, remembering that $P(c_{b})=1$.
If we also define an \emph{experimental centrality}, $c_{x}$, as
\begin{equation}
c_{x}\equiv\intop_{x}^{\infty}P(X)\,\mathrm{d}X\label{eq:exp-centrality}
\end{equation}
\emph{i.e.} the fraction of all events with $X\geq x$, then we can
also rewrite the denominator, giving finally for the centrality distribution
corresponding to our event selection

\begin{equation}
P(c_{b}|x_{1}\leq X\leq x_{2})=\frac{\int_{x_{1}}^{x_{2}}P(X|c_{b})\:\mathrm{d}X}{c_{x_{1}}-c_{x_{2}}}\label{eq:p_cb_for_x_cuts}
\end{equation}

More generally, events may be selected in many different ways, not
only using simple cuts, and not necessarily using the same observable
$X$ as that which is used for centrality estimation: for a generic
experimental sample $\mathbb{S}$ the centrality distribution will
be given by
\begin{equation}
P(c_{b}|\mathbb{S})=\frac{\int P(X|c_{b})\frac{P(X|\mathbb{S})}{P(X)}\,\mathrm{d}X}{\int P(X|\mathbb{S})\,\mathrm{d}X}\label{eq:general-cent-dist-for-selection}
\end{equation}
where $P(X|\mathbb{S})$ is the sample distribution of $X$ (\emph{i.e.}
a histogram of $X$ filled from the events in the sample), and the
integrals are over the full domain of $X$. Therefore once $P(X|c_{b})$
has been determined by fitting the experimental $P(X)$ distribution,
it can be used to estimate the centrality distribution for any given
selection of data.

Finally, absolute impact parameter distributions can be deduced from
the calculated centrality distributions by a suitable change of variables:
\begin{equation}
P(b|\mathbb{S})=P(b)P(c_{b}(b)|\mathbb{S})\label{eq:general-cent-dist-to-ip-dist-transform}
\end{equation}
It should be noted that in this case it is necessary to assume a specific
form for the inclusive impact parameter distribution $P(b)$ and calculate
the corresponding relationship between $c_{b}$ and $b$, $c_{b}(b)$,
using Eq. \ref{eq:b-centrality}.

\subsection{Specific implementation\label{subsec:Specific-implementation}}

In order to apply the method to data, specific choices and approximations
must be made. These concern the p.d.f. $f$ to use in Eq. \ref{eq:fluctuation-kernel},
the parametrization of the centrality dependence of the mean, $\overline{X}$,
and the centrality dependence of the reduced fluctuation, $\theta$.

\subsubsection{Fluctuation kernel, $f$\label{subsec:Fluctuation-kernel,}}

As observables most strongly correlated with impact parameter in heavy-ion
collisions can be assumed to result from a sum of many independent
microscopic processes (\emph{e.g.} nucleon-nucleon collisions), a
natural choice would be a gaussian or normal distribution \citep{Das2018Relating}.
However, as most observables related to collision violence only take
positive values, $X\geq0$, the gaussian distribution has the disadvantage
for small $\overline{X}$ that negative values may occur with finite
probability. Therefore, as shown in \citep{Rogly2018Reconstructing},
a better choice is the gamma distribution
\begin{equation}
f[k,\theta]=\frac{1}{\Gamma(k)\theta^{k}}X^{k-1}\mathrm{e}^{-X/\theta}\label{eq:gamma_distribution}
\end{equation}
with $k=\overline{X}/\theta$. The gamma distribution is only defined
for $X\geq0$, being asymmetric for small $\overline{X}$ and tending
towards a gaussian distribution asymptotically for large $\overline{X}$.

Another criterion for the choice of $f$ is the nature of the observable
$X$: is it a continuous or a discrete variable? Although in principle
the gaussian or gamma distributions can be used for both, in the case
of discrete variables such as particle multiplicities, one might think
that a discrete probability distribution may be more suitable. However,
the most likely candidates, the Poisson distribution, the binomial
distribution and the negative binomial distribution, each impose different
constraints on the reduced variance $\theta$: $\theta<1$ (binomial),
$\theta=1$ (Poisson) or $\theta>1$ (negative binomial). As will
be seen below, the experimental $P(X)$ distributions do not contain
sufficient information to constrain the dependency of \emph{both}
$\overline{X}$ \emph{and} $\theta$ on $c_{b}$, therefore rather
than constrain \emph{a priori} the value of $\theta$ by the choice
of $f$, it is preferable to use the gamma fluctuation kernel for
which $\theta$ can vary freely, even for observables which only take
discrete values.

\begin{figure*}
\includegraphics[clip,width=0.9\columnwidth]{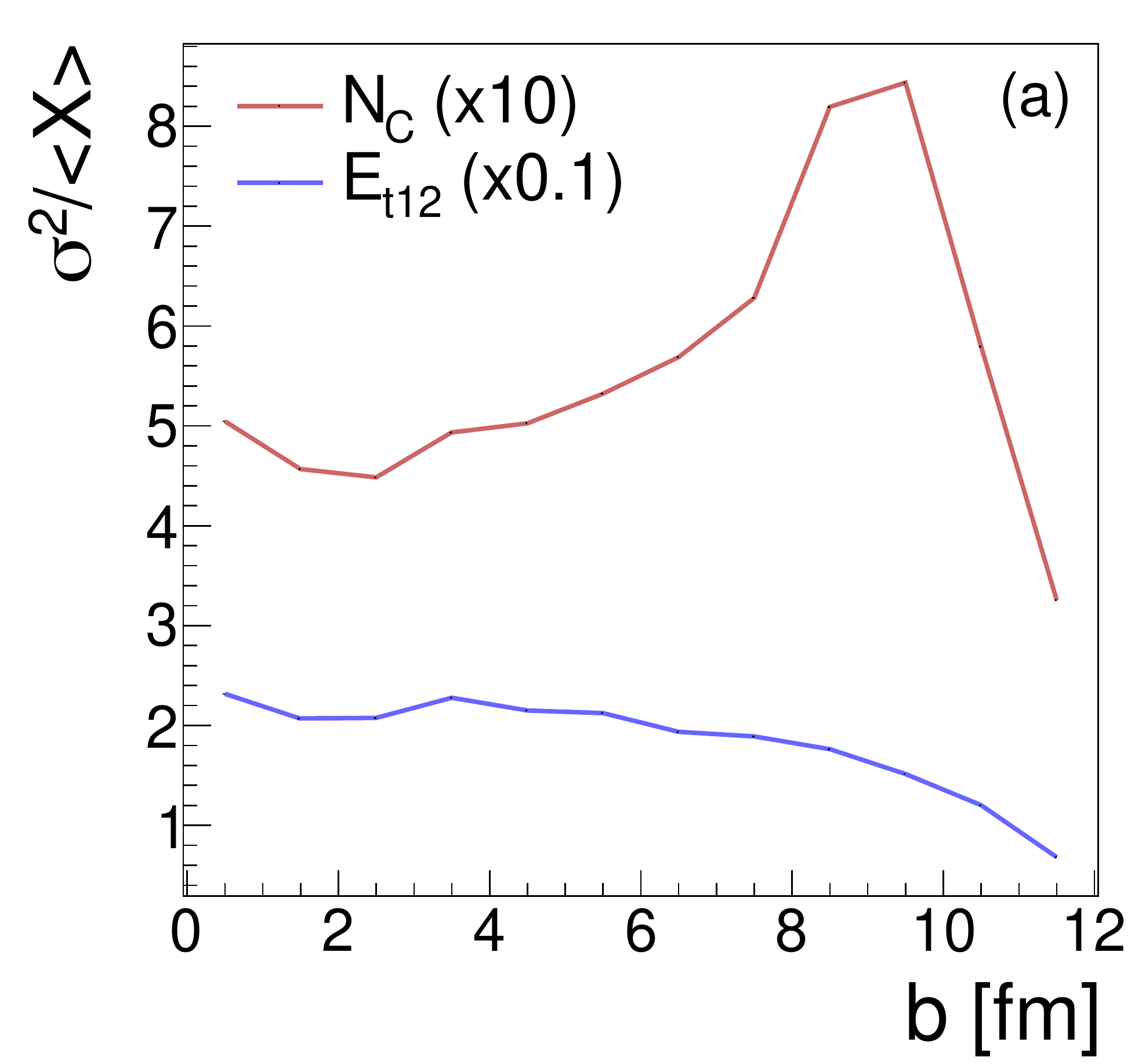}\hspace{1cm}\includegraphics[width=0.9\columnwidth]{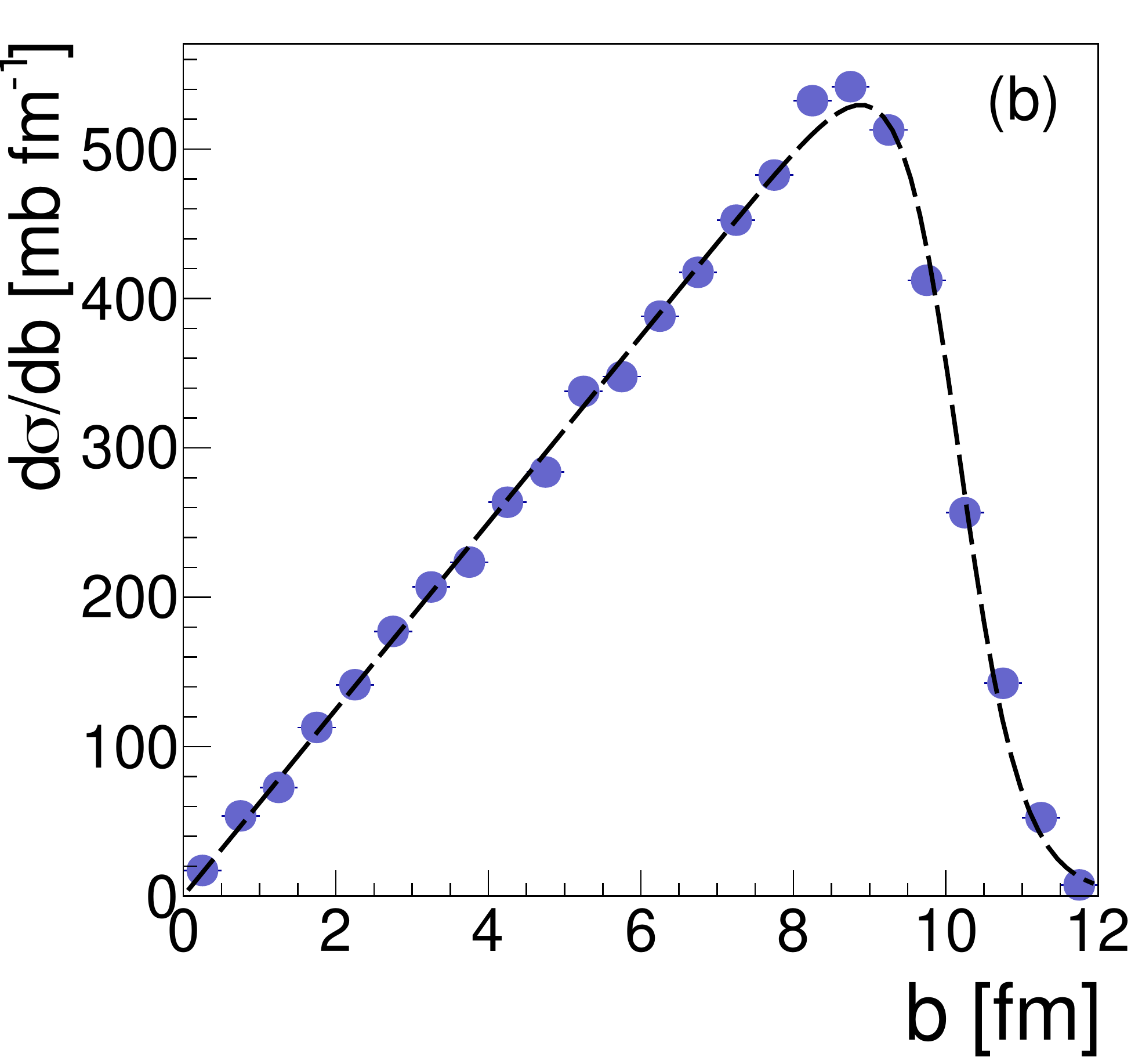}

\caption{AMD+GEMINI++ simulations of \collision{Ni}{58}{Ni}{58}{52} filtered
according to the acceptance of the INDRA array (see text for details).
(a) Reduced variances of the observables \mtot{} and \etlcp{} as
a function of impact parameter, $b$. (b) Symbols: differential cross-section
distribution of all detected simulated events as a function of $b$.
Dashed curve: fit to the distribution using Eq. \ref{eq:realistic-b-dist}
with $b_{0}=10.1$ fm and $\Delta b=0.4$ fm.\label{fig:Impact-parameter-dependence}}
\end{figure*}

\subsubsection{Centrality dependence of the mean, $\overline{X}$}

In order to reconstruct centrality distributions from experimental
$X$ distributions, the relationship between $\overline{X}$ and centrality
must be monotonic. We have chosen a functional form whose parameters
can vary freely while guaranteeing monotonicity and which is sufficiently
general to describe the typical shapes of such relationships as predicted
by various transport model calculations in this energy range (see
for example \citep{Nebauer1999Multifragmentation,Plagnol1999Onset,Zbiri2007,LeFevre2009,Bonnet2014,Li2018}). 

In the framework of the gamma distribution of Eq. \ref{eq:gamma_distribution},
it is in fact the parameter $k$ whose centrality dependence has to
be parameterized, for which we have used
\begin{equation}
k(c_{b})=k_{\mathrm{max}}\left[1-c_{b}^{\alpha}\right]^{\gamma}+k_{\mathrm{min}}\label{eq:my-k_cb}
\end{equation}
Note that this form corresponds to a monotonically decreasing function
of centrality (as expected for variables which increase with collision
violence): if a monotonically increasing function of centrality is
required, one can replace $c_{b}$ by $1-c_{b}$ in Eq. \ref{eq:my-k_cb}.
Only 4 free parameters are required. $\alpha$ and $\gamma$ can be
directly linked to the shape of the impact parameter dependence of
$\overline{X}$, making interpretation of fit results more immediate:
the value of $\alpha$ determines whether or not the observable's
evolution with $b$ presents a plateau for the most central collisions,
\emph{i.e.} when $\alpha\geq1$ there exists a range of small impact
parameters for which the derivative $\mathrm{d}k/\mathrm{d}b\approx0$,
which implies a lower limit to the observable's sensitivity to variations
of $b$ \textemdash{} the larger the value of $\alpha$, the larger
the range; the $\gamma$-parameter determines the concavity of the
curve \textemdash{} values of $\gamma>1$ lead to S-shaped curves
with an asymptotically zero derivative at $c_{b}=1$. 

$k_{\mathrm{max}}$ and $k_{\mathrm{min}}$ determine the maximum
mean value of the observable achieved in head-on collisions, i.e.
for $c_{b}=b=0$: 
\begin{equation}
X_{\mathrm{max}}=\overline{X}(b=0)=\theta(k_{\mathrm{max}}+k_{\mathrm{min}})\label{eq:Xmax-equation}
\end{equation}
The `offset' parameter $k_{\mathrm{min}}$ is important because
we cannot always assume that $X_{\mathrm{min}}=\theta k_{\mathrm{min}}$
is zero for the most peripheral recorded collisions. This is especially
clear when considering $X=N_{C}$, the total multiplicity of charged
products. All the data analyzed in this article were obtained with
an online DAQ trigger corresponding to a minimum number of fired telescopes
of between 3 and 5 depending on the system studied. For the lightest
systems considered in our study, the maximum charged particle multiplicity
can be as small as 20; in this case the role of $k_{\mathrm{min}}$
is far from negligible.

\subsubsection{Centrality dependence of reduced fluctuations, $\theta$\label{subsec:Centrality-dependence-of}}

As first pointed out in \citep{Das2018Relating}, the experimental
$P(X)$ distributions do not contain sufficient information to constrain
the dependency of \emph{both} $\overline{X}$ \emph{and} $\theta$
on $c_{b}$. The problem is under-constrained, in the sense that one
cannot extract two unknown functions, $\overline{X}(c_{b})$ and $\theta(c_{b})$,
from a single distribution, $P(X)$. Nevertheless, the fluctuations
for central collisions, $\theta(b=0)$, can be well-constrained by
the tail of the $P(X)$ distribution, which is dominated by fluctuations
around the mean value for $b\sim0$, $\overline{X}(b=0$). The rest
of the distribution gets contributions from many different impact
parameters so that fluctuations are averaged over, and $P(X)$ only
contains information about $\overline{X}(c_{b})$ away from the tail.
Therefore, as in \citep{Das2018Relating,Rogly2018Reconstructing},
we make the approximation that the reduced fluctuation parameter $\theta$
is constant for all centralities, and equal to its value for $b=0$:
in other words the variance of $X$ is proportional to its mean value
for all $b$, with a constant of proportionality $\theta$ which is
a free parameter of the fits, constrained by the tail of the $P(X)$
distributions.

\begin{figure*}
\includegraphics[width=0.9\columnwidth]{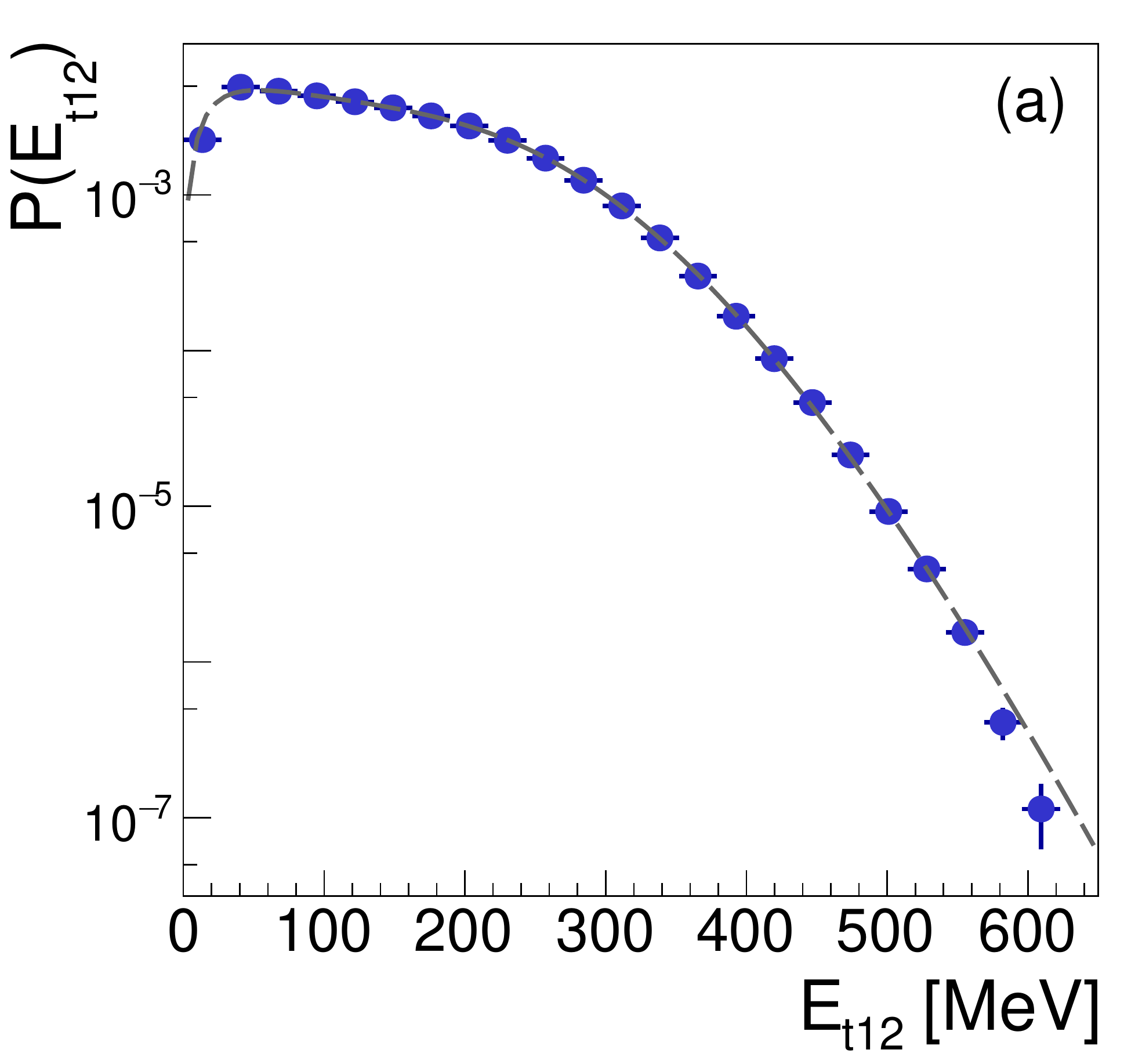}\hspace{1cm}\includegraphics[width=0.9\columnwidth]{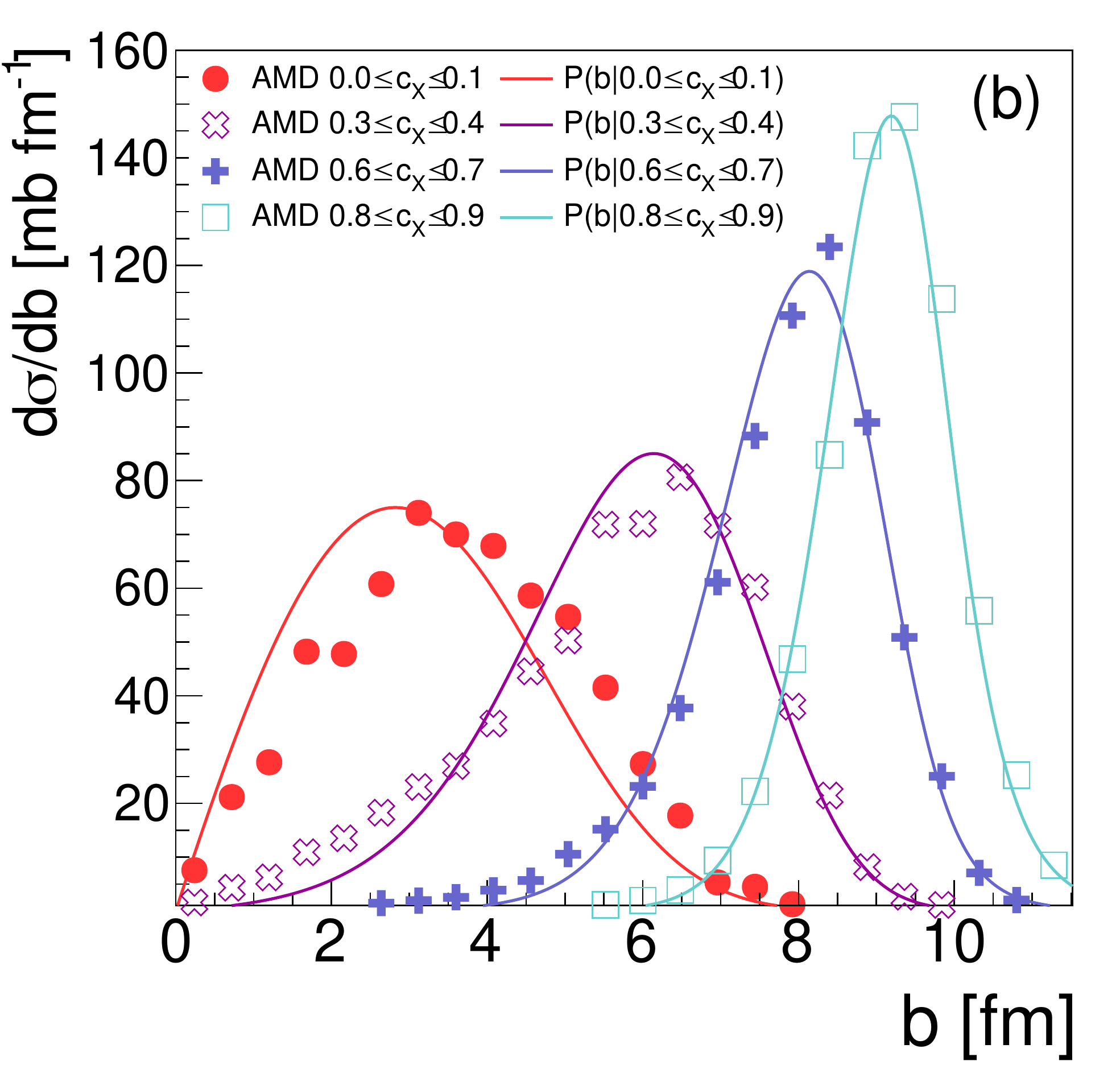}

\caption{Impact parameter distribution reconstruction for the observable \etlcp{}
for simulated data from \collision{Ni}{58}{Ni}{58}{52} (see text
for details).\label{fig:Fits-to-simulated} (a) Symbols: inclusive
probability distribution $P(E_{t12})$ of total transverse energy
of light charged particles. Statistical uncertainties are represented
by vertical bars when larger than the symbols. Dashed curve: fit using
Eq. \ref{eq:int-over-cb-p(x)} and the parametrizations of Eqs. \ref{eq:gamma_distribution},\ref{eq:my-k_cb}.
(b) Reconstructed impact parameter distributions (curves) calculated
using Eqs. \ref{eq:p_cb_for_x_cuts}, \ref{eq:general-cent-dist-to-ip-dist-transform},
compared to the actual distributions (symbols) for 4 different bins
of experimental centrality, $c_{X}$.}
\end{figure*}

\subsection{Test with a transport model calculation\label{subsec:Application-to-simulated}}

We have used the microscopic transport model AMD (Antisymmetrised
Molecular Dynamics, \citep{Ono2013,Ono2004a}) in order to simulate
a typical reaction from the INDRA dataset, \collision{Ni}{58}{Ni}{58}{52}.
The transport model calculation was stopped at the end of the dynamical
phase of the reaction, after 300 fm/c, and the resulting set of excited
primary nuclei was used as input to the statistical decay code GEMINI++
\citep{Charity2010Systematic,Mancusi2010Unified}. The detection of
the final products of these collisions by the INDRA array \citep{jdfrankland:Pouthas1995INDRA}
was then simulated using a software replica of the apparatus \citep{KaliVeda},
including a minimum-bias acquisition trigger requiring simultaneous
detection of 4 charged products or more (as during the experimental
measurement). Our goal here is not to test the AMD model, nor to compare
its predictions with data, but rather to have at our disposition a
set of simulated data which correctly reproduce at least the main
features of the experimental data, while linking the impact parameter
of each collision with its outcome based on the pertinent microscopic
physics ingredients for heavy-ion collisions in this energy range.

A first result of these calculations is shown in Figure \ref{fig:Impact-parameter-dependence}(a).
The reduced variance of the two observables which we will use in the
following, the total charged particle multiplicity, \mtot{}, and
the total transverse energy of light charged particles with $Z\leq2$,
\etlcp{}, is presented as a function of impact parameter. The variances
presented here include several sources of fluctuations: fluctuations
in the microscopic collision dynamics, fluctuations in the secondary
decay of excited primary fragments, and fluctuations due to the finite
acceptance of the INDRA array and other detection effects. It can
be seen that for all but the most peripheral collisions ($b<8$ fm),
the reduced variance of both observables can be considered approximately
constant. Therefore, the approximation of \ref{subsec:Centrality-dependence-of},
whereby the reduced fluctuation parameter $\theta$ is assumed to
be independent of the impact parameter, can be seen to have some justification,
at least as far as this modelization is concerned.

Figure \ref{fig:Fits-to-simulated}(a) shows an example of the quality
of fits to the inclusive data histograms which can be achieved using
Eq. \ref{eq:int-over-cb-p(x)} and the parametrizations of Eqs. \ref{eq:gamma_distribution},\ref{eq:my-k_cb},
here for the total transverse energy of light charged particles, \etlcp.
The parameter values deduced from this fit completely determine the
conditional probability distribution $P(E_{t12}|c_{b})$, and can
therefore be used to calculate centrality distributions for any arbitrary
selection of data. However, to calculate \emph{impact parameter} distributions,
as pointed out at the end of Sec. \ref{subsec:General-approach},
requires the knowledge of, or at least a good approximation to, the
full impact parameter distribution $P(b)$. In the present case, of
course, this distribution is accessible, as shown in Figure \ref{fig:Impact-parameter-dependence}(b).
Initially, the AMD calculations were run with a geometric impact parameter
distribution,
\[
P(b)=2\pi b
\]
for all $b\leq12$ fm. Let us recall that the equivalent of the experimental
DAQ trigger in this simulation is the detection of at least 4 \emph{charged}
particles in coincidence: as can be seen, for the most peripheral
collisions ($b>9$ fm) the probability of such events decreases to
zero. Such a distribution can be well-fitted by $P(b)=2\pi bP_{R}(b)$
with
\begin{equation}
P_{R}(b)=\frac{1}{1+\exp\left(\frac{b-b_{0}}{\Delta b}\right)}\label{eq:realistic-b-dist}
\end{equation}
as shown by the dashed curve in Figure \ref{fig:Impact-parameter-dependence}(b),
for parameter values $b_{0}=10.1$ fm and $\Delta b=0.4$ fm. Similar
distributions have been observed previously using different models
applied to different reaction systems of the INDRA dataset (see for
example Figure 2(a) of \citep{Vient2018a}), and suggest that in general
for ``minimum-bias'' INDRA data, Eq. \ref{eq:realistic-b-dist}
with $\Delta b\approx0.4$ fm gives a good approximation to the full
impact parameter distribution. The formulae allowing to calculate
$b_{0}$ for a given value of $\Delta b$ knowing the total measured
cross-section, and the relation between centrality and impact parameter,
$c_{b}(b)$, needed for Eq. \ref{eq:general-cent-dist-to-ip-dist-transform},
are given in Appendix \ref{sec:Inclusive-impact-parameter}.

Examples of reconstructed impact parameter distributions calculated
using the deduced form of $P(E_{t12}|c_{b})$ are shown in Figure
\ref{fig:Fits-to-simulated}(b). Here we have gated the simulated
data according to 4 bins of experimental centrality, $c_{X}$, calculated
with the \etlcp{} variable (see Eq. \ref{eq:exp-centrality}). As
can be seen, the reconstructed distributions are very similar to the
actual impact parameter distributions which in this case can be simply
generated by applying the \etlcp{} cuts directly to the model. Both
mean impact parameter and width of the distribution for each centrality
bin are very well reproduced. For the most central bin ($c_{X}\leq0.1$,
corresponding to the upper 10\% of the \etlcp{} distribution), although
a small shift can be observed, leading to a slight underestimation
of the mean impact parameter for these events, the reconstructed distribution
is still a very good approximation to the actual one.

If this new method underestimates slightly the impact parameters for
a very central cut, it is as nothing compared to the underestimation
that would result from the commonly-used sharp cut-off approximation
(SCA). Within the SCA, the cross-section associated with any selection
of ``the most central collisions'' is assumed to correspond to a
geometric (triangular) distribution of impact parameters between $b=0$
and $b=b_{\mathrm{cut}}$, where $b_{\mathrm{cut}}$ is the assumed
upper limit of impact parameters retained by the cut. In this case,
the fraction of all detected events retained by the cut, i.e. the
upper limit of the experimental centrality $c_{X}$, is simply related
to the upper limit of the impact parameter by $c_{X}=(b_{\mathrm{cut}}/b_{\mathrm{max}})^{2}$.
$b_{\mathrm{max}}$ is the SCA upper limit of impact parameter for
all detected events, which in the present case is to a very good approximation
equivalent to the value of $b_{0}$ in Eq. \ref{eq:realistic-b-dist}.
Thus the SCA predicts for a centrality cut $c_{X}\leq0.1$ an upper
limit of impact parameter $b_{\mathrm{cut}}\approx3$ fm, whereas
in reality the impact parameters of this selection can be seen in
Figure \ref{fig:Fits-to-simulated}(b) to have a very broad distribution
which extends up to 8 fm.

\begin{table}
\begin{tabular}{ccr@{\extracolsep{0pt}.}lcr}
System & Mass & \multicolumn{2}{c}{$E_{\mathrm{proj}}$} & Trigger & Events\tabularnewline
 & asymmetry & \multicolumn{2}{c}{{[}MeV/} & multiplicity & \tabularnewline
 &  & \multicolumn{2}{c}{nucleon{]}} &  & \tabularnewline
\hline 
\multirow{4}{*}{\col{Ar}{36}{KCl}{} \citep{Metivier2000Mass}} & \multirow{4}{*}{0.00} & 31&54 & 3 & 3216332\tabularnewline
 &  & 39&97 & 3 & 3496188\tabularnewline
 &  & 51&66 & 3 & 2391311\tabularnewline
 &  & 74&00 & 3 & 3337570\tabularnewline
\hline 
\multirow{7}{*}{\col{Ar}{36}{Ni}{58} \citep{Bacri1995Onset,Rivet1996Vaporization}} & \multirow{7}{*}{0.23} & 31&54 & 3 & 8259867\tabularnewline
 &  & 39&97 & 3 & 7234383\tabularnewline
 &  & 51&66 & 3 & 8599855\tabularnewline
 &  & 63&03 & 3 & 5020363\tabularnewline
 &  & 74&00 & 4 & 7648474\tabularnewline
 &  & 83&63 & 4 & 4657028\tabularnewline
 &  & 95&22 & 4 & 9799670\tabularnewline
\hline 
\multirow{6}{*}{\col{Ni}{58}{Ni}{58} \citep{Cussol2002,Galichet2009Isospin}} & \multirow{6}{*}{0.00} & 31&98 & 4 & 4538513\tabularnewline
 &  & 52&00 & 4 & 4738429\tabularnewline
 &  & 63&63 & 4 & 4473639\tabularnewline
 &  & 73&96 & 4 & 5198692\tabularnewline
 &  & 82&00 & 4 & 5578566\tabularnewline
 &  & 90&00 & 4 & 9144521\tabularnewline
\hline 
\multirow{5}{*}{\col{Ni}{58}{Au}{197} \citep{jdfrankland:Bellaize2002Multifragmentation}} & \multirow{5}{*}{0.55} & 31&98 & 4 & 7448285\tabularnewline
 &  & 52&00 & 4 & 7941858\tabularnewline
 &  & 63&63 & 4 & 4720169\tabularnewline
 &  & 73&96 & 4 & 6685519\tabularnewline
 &  & 82&00 & 4 & 7398023\tabularnewline
\hline 
\multirow{5}{*}{\xesn \citep{Marie1997Hot,Plagnol1999Onset}} & \multirow{5}{*}{0.04} & 24&98 & 4 & 5288164\tabularnewline
 &  & 32&00 & 4 & 3916797\tabularnewline
 &  & 38&98 & 4 & 5261377\tabularnewline
 &  & 45&00 & 4 & 6067739\tabularnewline
 &  & 50&13 & 4 & 5792220\tabularnewline
\hline 
\multirow{3}{*}{\xesngsi({*}) \citep{Frankland2005}} & \multirow{3}{*}{0.02} & 65&00 & 3 & 881642\tabularnewline
 &  & 80&00 & 3 & 424357\tabularnewline
 &  & 100&00 & 3 & 1328486\tabularnewline
\hline 
\multirow{4}{*}{\col{Au}{197}{Au}{197}({*}) \citep{ukasik2005}} & \multirow{4}{*}{0.00} & 40&00 & 3 & 2783629\tabularnewline
 &  & 60&00 & 3 & 7589902\tabularnewline
 &  & 80&00 & 3 & 3545170\tabularnewline
 &  & 100&00 & 3 & 10691556\tabularnewline
\end{tabular}

\caption{Characteristics of collisions studied in this work: mass asymmetry
$|A_{p}-A_{t}|/(A_{p}+A_{t})$, beam energy, DAQ trigger multiplicity
and total number of recorded events. References are given to the original
papers where details of the data-taking can be found. Systems marked
with an asterisk were measured at GSI, all others at GANIL. \label{tab:Characteristics-of-systems}}
\end{table}

\section{Results \label{sec:Results-for-INDRA}}

In the following we will present the results of applying the method
presented in Sec. \ref{sec:Method} to data for a wide range of different
colliding systems measured with INDRA, which are summarized in Table
\ref{tab:Characteristics-of-systems}. The data concern the two observables
which are most commonly used for centrality estimation and/or selections
in this energy range, namely the total multiplicity of charged reaction
products, \mtot, and the total transverse energy of light charged
particles (LCP, isotopes of $Z=1,2$ nuclei), \etlcp. \mtot\  is
the impact parameter filter most commonly-used by many different groups
in the intermediate energy range, while \etlcp\  has been especially
used by the INDRA collaboration as it exploits the very high, angle-
and centrality-independent efficiency of the array for detection of
LCP.

\subsection{Experimental details\label{subsec:Experimental-details}}

Impact parameter estimation and sorting of HIC in the energy range
\amev{20\textendash 100} requires powerful multidetector arrays with
high granularity and $4\pi$ angular coverage. Let us briefly recall
here that INDRA \citep{jdfrankland:Pouthas1995INDRA,Pouthas1996Electronics}
is one of a second generation of $4\pi$ charged particle arrays,
in continued use for the study of HIC at GANIL in Caen (and briefly
at GSI, Darmstadt) since 1993. Its 336 multi-layer detection modules
covering $90\%$ of the solid angle around the target, low detection
and identification thresholds, and minimum-bias trigger logic based
on the number of fired modules make it ideally suited for impact parameter
reconstruction in this energy range. References to the original papers
where details of the data-taking can be found for each measured reaction
are given in Table \ref{tab:Characteristics-of-systems}.

Also in Table \ref{tab:Characteristics-of-systems}, as well as the
mass asymmetry, projectile energy and number of recorded events, we
give also the trigger multiplicity (corresponding to the minimum number
of fired modules required to record an event, which may include $\gamma$-ray,
electron, pion or neutron detection in the CsI scintillators) for
each reaction. In the offline analysis the same condition was applied
to the reconstructed events (corresponding to a minimum number of
correctly identified charged products, thus excluding $\gamma$-rays
etc.).

\subsection{Results of fits to data\label{subsec:Results-of-fits}}

\begin{figure*}
\includegraphics[width=0.9\columnwidth]{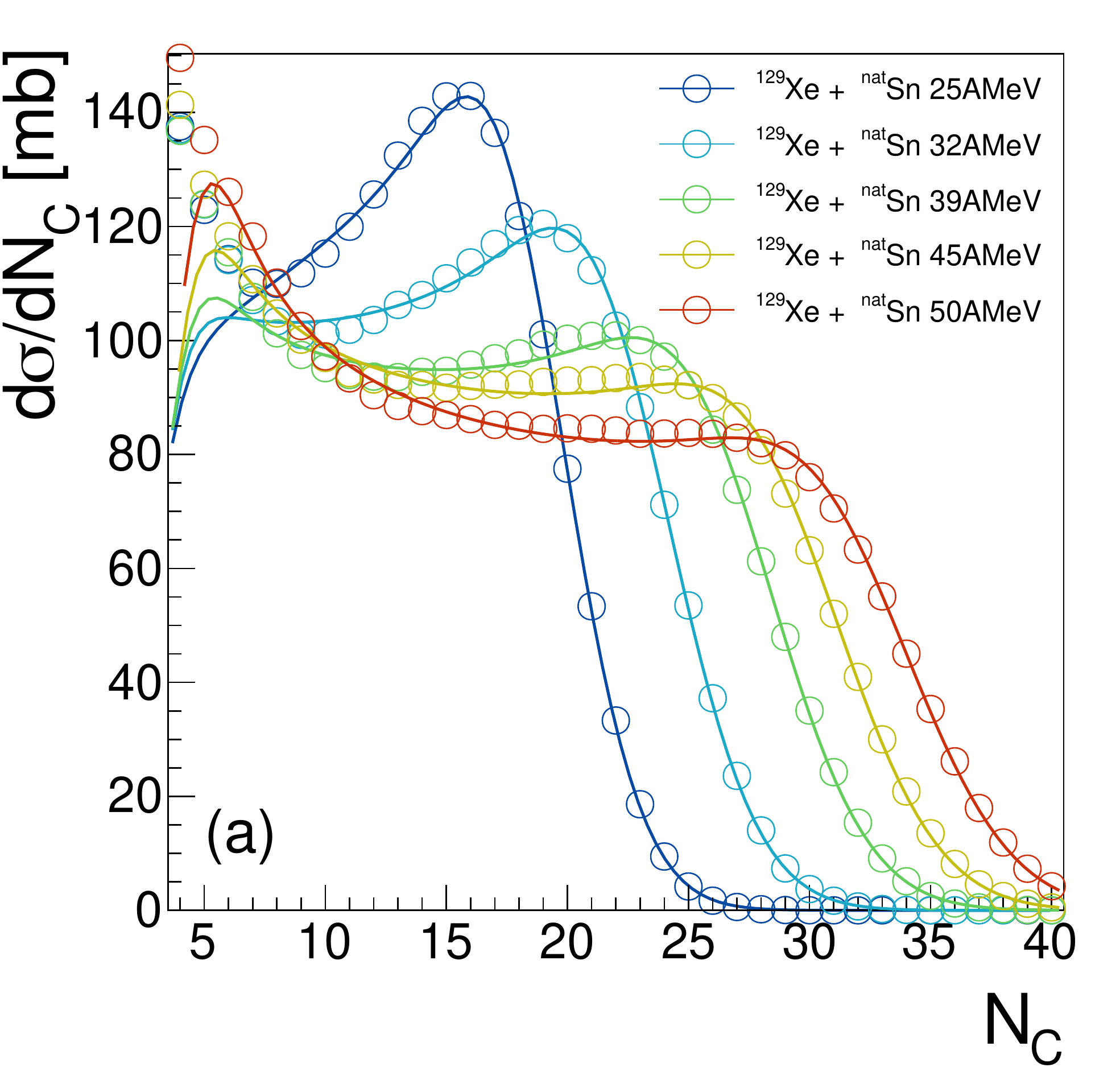}\hspace{1cm}\includegraphics[width=0.9\columnwidth]{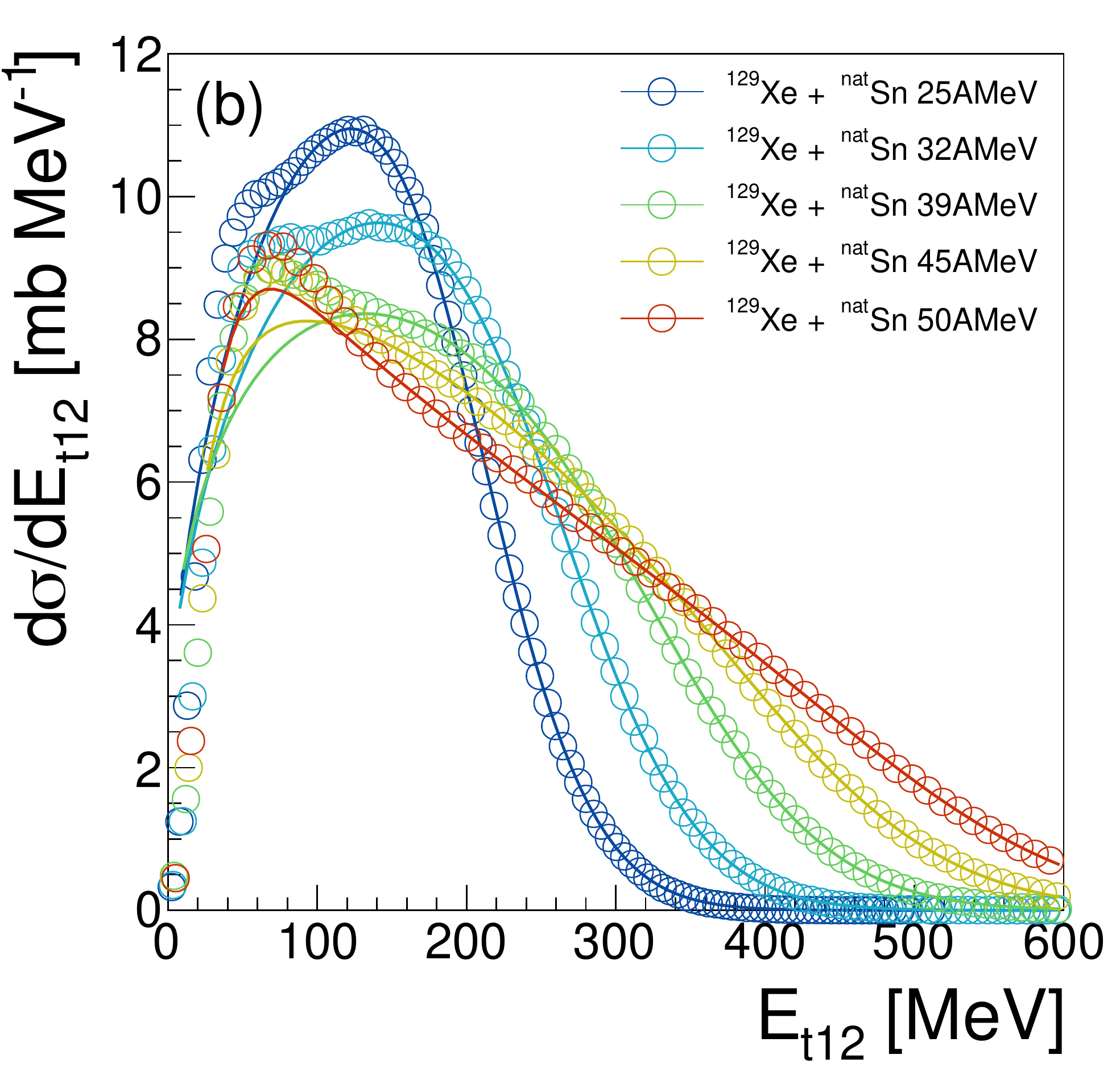}

\includegraphics[width=0.9\columnwidth]{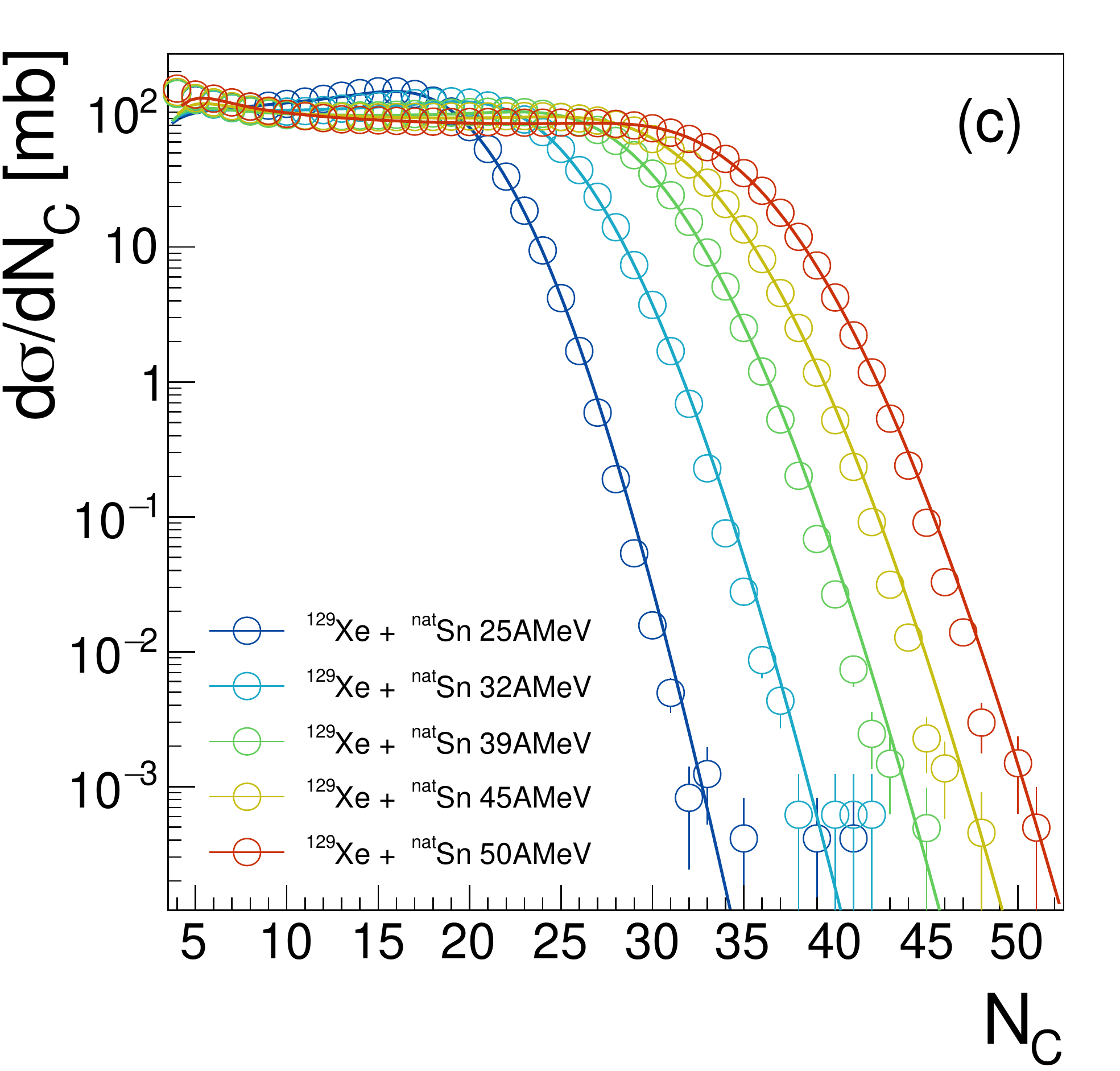}\hspace{1cm}\includegraphics[width=0.9\columnwidth]{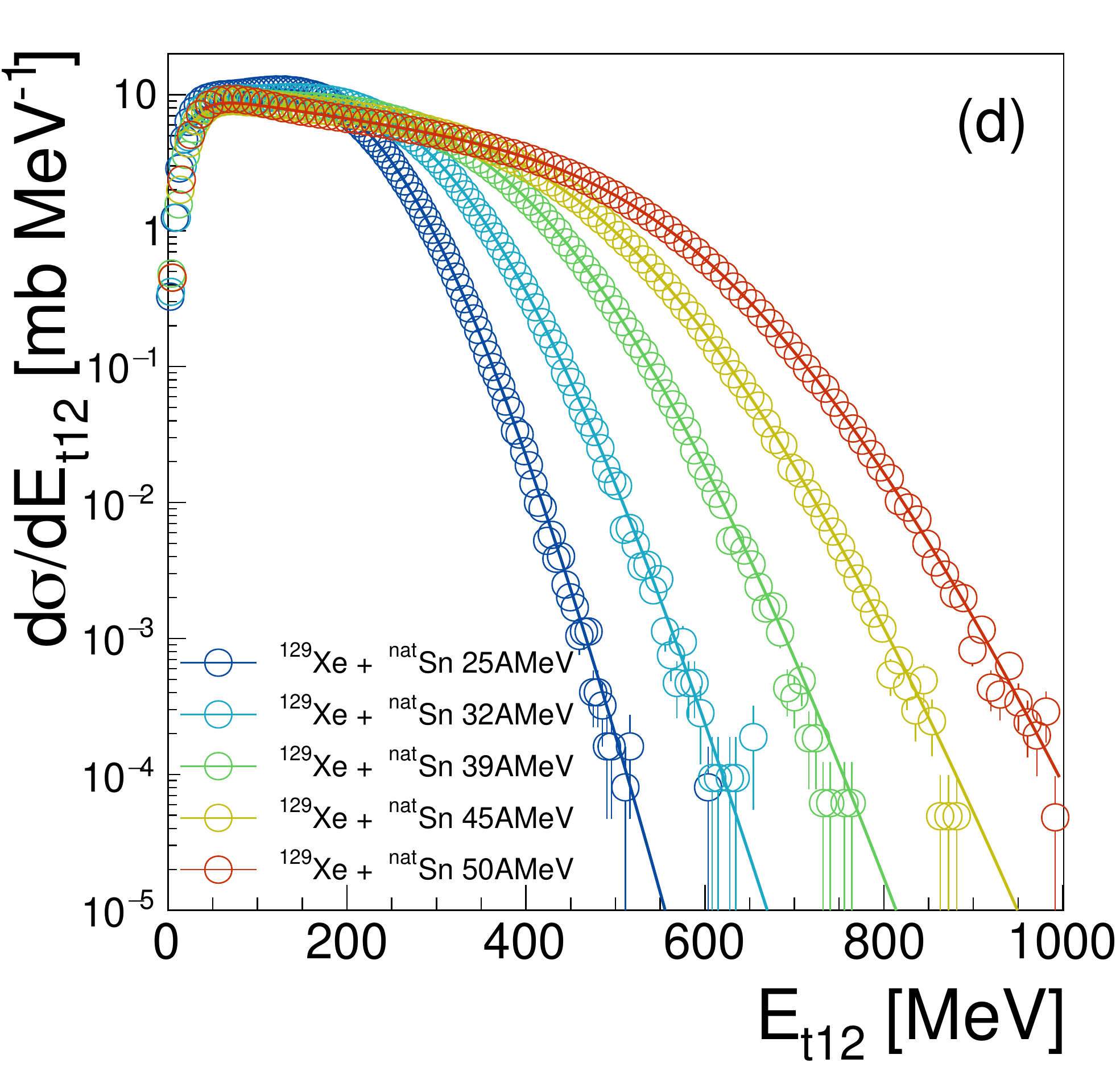}

\caption{Results of fits to the inclusive distributions of \mtot\  (left column,
(a) \& (c)) and \etlcp\  (right column, (b) \& (d)) for the \xesn\ 
data. Each distribution is presented with both linear (top row, (a)
\& (b)) and logarithmic (bottom row, (c) \& (d)) $y$-axes. Statistical
uncertainties on the data are shown when not smaller than the symbols.\label{fig:fits-xesn-mtot-et12}}
\end{figure*}

Examples of fits to the inclusive distributions of the observables
\mtot\  and \etlcp\  are presented in Figure \ref{fig:fits-xesn-mtot-et12},
for the $M\geq4$ \xesn\  data. Fits were performed using Eqs. \ref{eq:int-over-cb-p(x)},
\ref{eq:gamma_distribution} and \ref{eq:my-k_cb} and fit parameters
are given in Tables \ref{tab:multiplicity-fits} and \ref{tab:et12-fits}
for \mtot{} and \etlcp{}, respectively. Using the measured reaction
cross-sections for this data \citep{Plagnol1999Onset}, the experimental
and fitted $P(X)$ distributions are presented here as differential
cross-sections. To better appreciate the quality of the fits, for
both low and high statistics regions of the distributions, each is
presented with both linear (top row) and logarithmic (bottom row)
$y$-axes. 

Apart from the lowest \mtot\  or \etlcp\  values, close to the trigger
threshold, the shapes of the experimental distributions are globally
well reproduced by each fit, especially the exponential tails for
the highest multiplicities/energies. Reduced $\chi^{2}$ values for
each fit are reported in Tables \ref{tab:multiplicity-fits} and \ref{tab:et12-fits}.
For \etlcp\  this goodness-of-fit parameter is generally excellent
($\chi^{2}\sim1$), whereas for \mtot\  the values are less satisfactory,
despite the visual impression of adequate fits: this is due to the
sharp decrease of the fitted distributions at small \mtot\ , which
due to the high statistics in this region dominates the overall $\chi^{2}$
values. Nevertheless, it can be remarked that the deduced $X_{\mathrm{min}}$
values for \mtot\  follow remarkably well the minimum multiplicity
imposed by the trigger, including the increase from $M\geq3$ to $M\geq4$
for the \arni\  data at \amev{74} (see Table \ref{tab:multiplicity-fits}).
Fits of similar quality for both observables were obtained for all
data in this study, the parameters of which are given in Tables \ref{tab:multiplicity-fits}
and \ref{tab:et12-fits}.

\subsection{Deduced probability distributions for impact parameter and observable}

\begin{figure}
\includegraphics[clip,width=1\columnwidth]{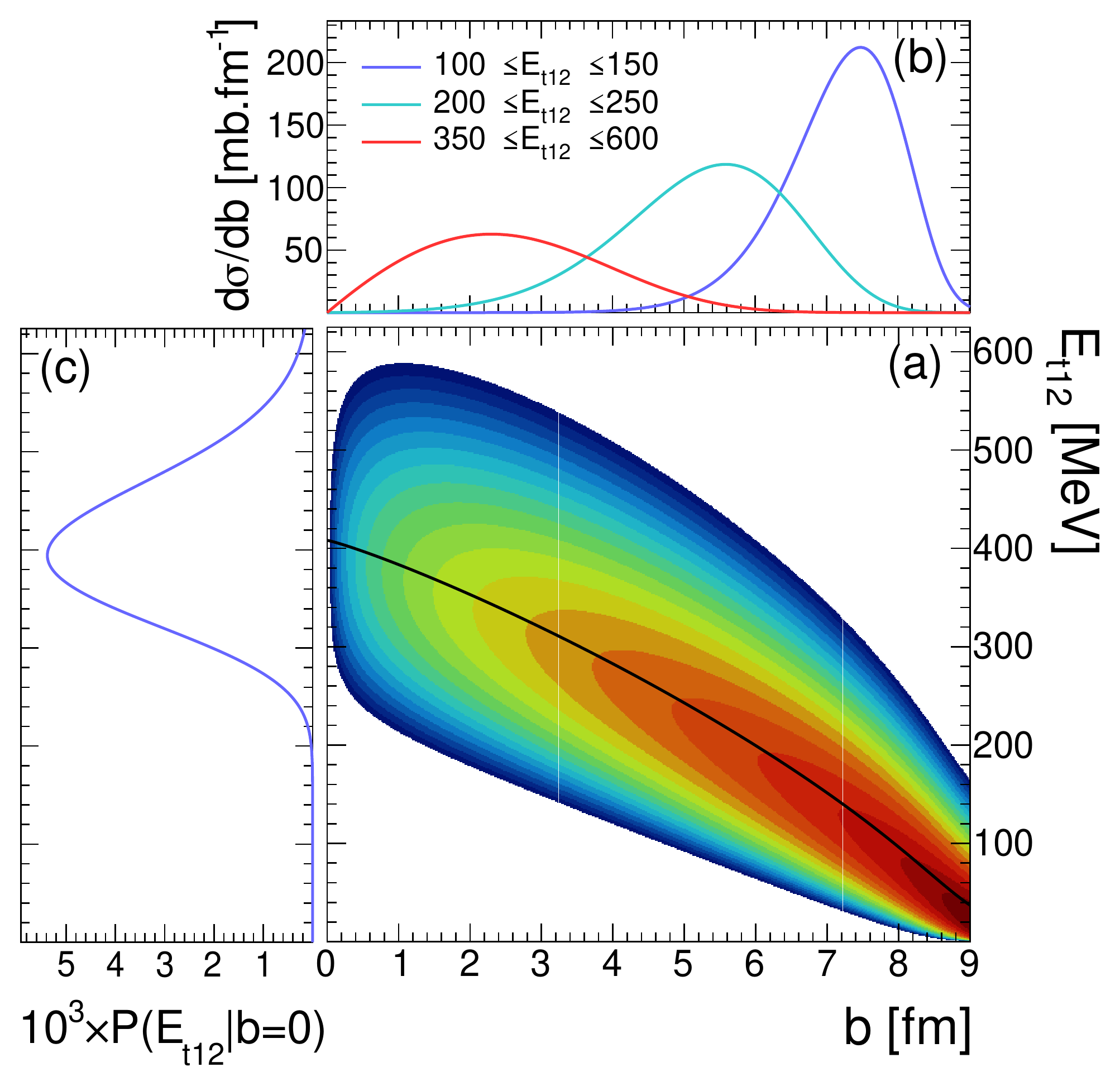}

\caption{Results obtained from the fit to the experimental \etlcp{} distribution
for \collision{Xe}{129}{Sn}{nat}{39}: (a) reconstructed joint probability
distribution $P(E_{\mathrm{t12}},b)$ (logarithmic color scale) and
the mean value $\overline{E_{\mathrm{t12}}}(b)$ (black curve); (b)
impact parameter distributions for three cuts in \etlcp{} indicated
in the legend; (c) distribution of \etlcp{} for $b=0$.\label{fig:Results-obtained-from-fit}}

\end{figure}

Fitting the experimental $P(X)$ distribution determines the parameters
of the conditional probability distribution $P(X|c_{b})$ or $P(X|b)$,
and hence the joint probability distribution of observable and impact
parameter, $P(X,b)=P(X|b)P(b)$. An example is shown in Figure \ref{fig:Results-obtained-from-fit}(a)
for the case of \collision{Xe}{129}{Sn}{nat}{39} using the total
transverse energy of light charged particles, \etlcp{}. As shown
in the figure, the joint probability distribution encodes not only
the dependence of the mean value \meanetlcp{} on impact parameter,
given by $\overline{X}=\theta k(c_{b})$ (see Eqs. \ref{eq:gamma_distribution},
\ref{eq:my-k_cb}), but crucially also the fluctuations around this
mean value at fixed $b$, determined by the parameter $\theta$.

It is clear from the joint probability distribution $P(E_{\mathrm{t12}},b)$
shown in Figure \ref{fig:Results-obtained-from-fit}(a) that any selection
of events using \etlcp{} cuts will necessarily lead to broad impact
parameter distributions. This is confirmed by Figure \ref{fig:Results-obtained-from-fit}(b)
which presents impact parameter distributions corresponding to three
$E_{\mathrm{t12}}$ bins, calculated according to Eqs. \ref{eq:p_cb_for_x_cuts}
and \ref{eq:general-cent-dist-to-ip-dist-transform}. From the figure
it can be seen that these $P(b|x_{\mathrm{1}}\leq E_{t12}\leq x_{\mathrm{2}})$
are simply projections of $P(E_{\mathrm{t12}},b)$ on to the impact
parameter axis, integrated over different ranges of \etlcp{}. The
distributions of $b$ for the three cuts are very broad and overlap
to a large extent.

Finally, Figure \ref{fig:Results-obtained-from-fit}(c) shows a specific
projection of $P(E_{\mathrm{t12}},b)$ on to the \etlcp{} axis, corresponding
to $P(E_{t12}|b=0)$, the deduced distribution of \etlcp{} for head-on
collisions. It is also, as expected, very broad: the variance of this
distribution is determined by the product of the reduced fluctuation
parameter, $\theta$, and the mean value $\overline{E_{t12}}(b=0)$
of the \etlcp{} observable for $b=0$ collisions. The values of $\theta$
for this system and all others, and for both \mtot{} and \etlcp{}
observables are given in Tables \ref{tab:multiplicity-fits} and \ref{tab:et12-fits}.
For the fits to \xesn{} data shown in Figure \ref{fig:fits-xesn-mtot-et12},
they are much smaller for \mtot{} ($\theta\sim0.3$) than for \etlcp{}
($11\leq\theta\leq16$). In fact, as a general result for the whole
set of reactions studied here, we find that total charged multiplicity
fluctuations are relatively small ($\theta<1$) and nearly independent
of bombarding energy, whereas \etlcp{} fluctuations are relatively
large ($\theta\gg1$) and increase with beam energy. Let us recall
that a similar difference in order of magnitude for $\theta$ was
also found with the AMD calculations shown in Figure \ref{fig:Impact-parameter-dependence}(a). 

\subsection{System mass- and bombarding-energy dependence of mean values of observables
for $b=0$}

\begin{figure}
\includegraphics[clip,width=1\columnwidth]{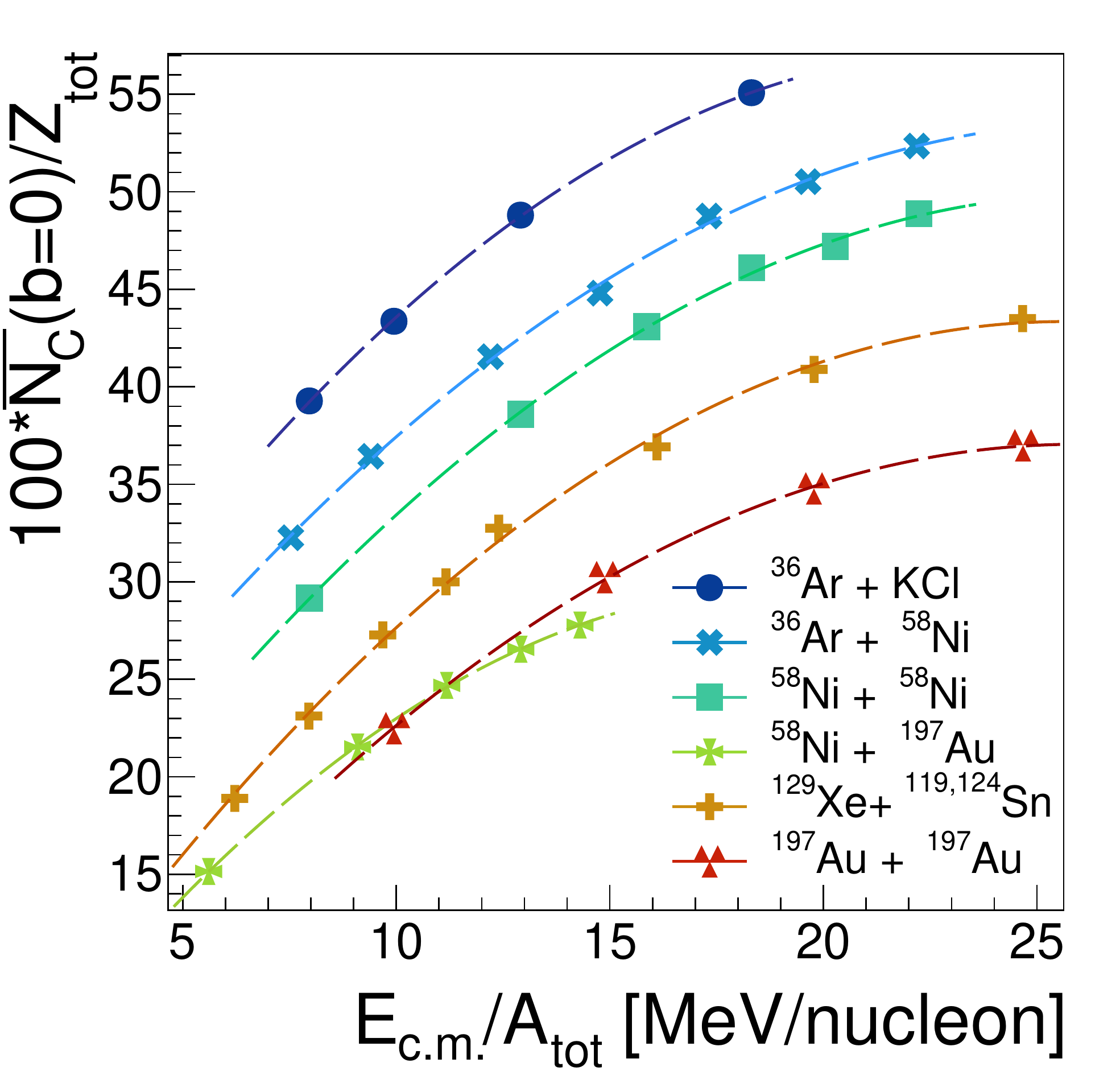}

\caption{Extrapolated mean values of total charged product multiplicity for
$b=0$ collisions deduced from fits, $\overline{N_{\mathrm{C}}}(b=0)$,
normalized to total system charge $Z_{\mathrm{tot}}$, as a function
of available center of mass energy per nucleon $E_{\mathrm{c.m.}}/A_{\mathrm{tot}}$
for all systems. Dashed curves are parabolic fits.\label{fig:extrapolated-mtot}}
\end{figure}

\begin{figure}
\includegraphics[width=1\columnwidth]{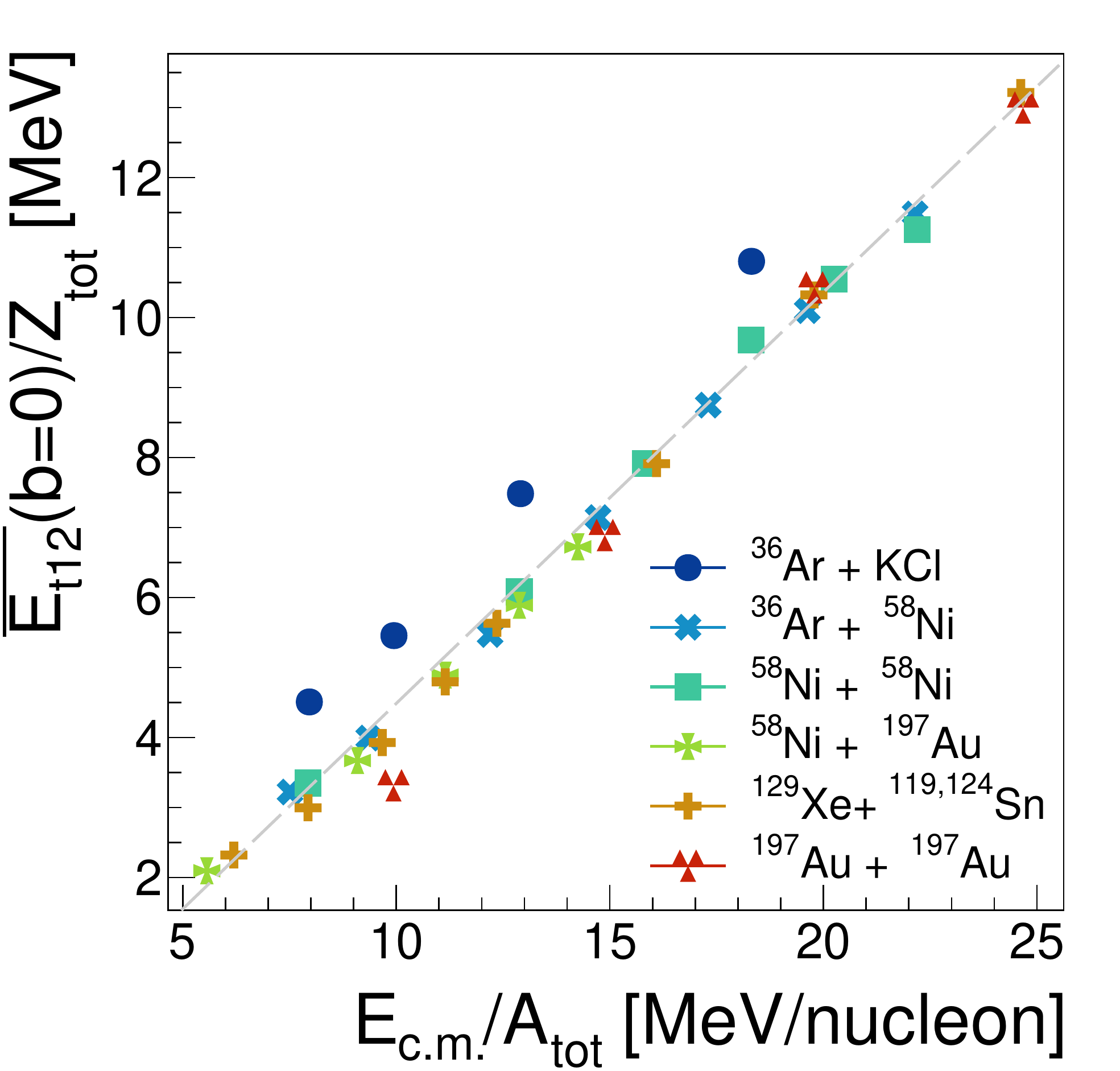}

\caption{Extrapolated mean values of total transverse energy of light charged
particles for $b=0$ collisions deduced from fits, $\overline{E_{\mathrm{t12}}}(b=0)$,
normalized to total system charge $Z_{\mathrm{tot}}$, as a function
of available center of mass energy per nucleon $E_{\mathrm{c.m.}}/A_{\mathrm{tot}}$
for all systems. The dashed gray line is a linear fit to all the data
except \arkcl{}.\label{fig:Extrapolated-etlcp}}
\end{figure}

Figures \ref{fig:extrapolated-mtot} and \ref{fig:Extrapolated-etlcp}
present the evolution of \meanmtot{} and \meanetlcp{} extrapolated
to head-on collisions for all studied reactions as a function of the
available center of mass energy, in order to allow equivalent comparison
for symmetric and asymmetric colliding systems. 

The total multiplicities in Figure \ref{fig:extrapolated-mtot} have
been normalized to the size of each system, $Z_{\mathrm{tot}}$, given
by the sum of projectile and target atomic numbers. $Z_{\mathrm{tot}}$
is the maximum possible charged product multiplicity for each reaction,
if complete disintegration into hydrogen isotopes and (undetected)
neutrons were to occur. Thus the quantity $\overline{N_{C}}(b=0)/Z_{\mathrm{tot}}$
reflects the degree to which complete disintegration would occur in
head-on collisions for each reaction. For all systems this ratio increases
in a similar fashion with available energy, closely following a parabolic
dependence (shown by the dashed curves in Figure \ref{fig:extrapolated-mtot}).
It should be noted that the \anyxesn{} data, taken in two separate
experimental campaigns, at GANIL (\xesn{}: $E_{\mathrm{proj}}\leq$\amev{50})
and at GSI (\xesngsi{}: $E_{\mathrm{proj}}\geq$\amev{65}), follows
a continuous evolution over the full energy range. There is also a
systematic system size/mass dependence of the ratio, which is larger
at a given available energy for smaller systems. Thus for $E_{\mathrm{c.m.}}/A_{\mathrm{tot}}\sim$\amev{20}
it can be seen that the \nini{} system reaches 50\% of complete disintegration,
whereas for \auau{} the degree is no greater than 35\%. We remark
that the asymmetric system, \niau{}, does not appear to follow the
same systematic trend: it could have been expected to have the same
dependence on $E_{\mathrm{c.m.}}/A_{\mathrm{tot}}$ as \anyxesn{},
which has approximately the same total mass and charge but a quasi-symmetric
entrance channel.

Figure \ref{fig:Extrapolated-etlcp} shows the extrapolated \meanetlcp{}
for $b=0$ collisions for all systems, again as a function of available
energy. The normalization $\overline{E_{t12}}(b=0)/Z_{\mathrm{tot}}$
to the total system charge here has no direct physical interpretation,
but empirically we have found that it leads to a near-universal scaling
of the data which collapse onto a linear dependence of the total transverse
energy of LCP that would be achieved in head-on collisions as a function
of the c.m. energy. Again, it can be remarked that data from campaigns
in both GANIL and GSI (\xesngsi{}: $E_{\mathrm{proj}}\geq$\amev{65}
and \auau{}) follow the same systematic trends, including this time
the \niau{} system; on the other hand, here it is the \arkcl{} data
which do not follow the trend, in actual fact because the unscaled
\meanetlcp{} values for this system are almost identical to those
for \arni{}.

We know of no reason why the mean transverse energy of light charged
particles \etlcp{} for $b=0$ collisions should scale in this way,
over such a wide range of system sizes, asymmetries, and bombarding
energies. It is our opinion that this result should be compared with
the predictions of different transport models in this energy range,
for which it could provide a benchmark test. 

\subsection{System mass- and bombarding-energy dependence of mean impact parameters
of events selected with a high-\etlcp{} cut}

As we recalled in the introduction, it is a well-known fact that when
trying to select more and more central collisions using high-\mtot{}
or \etlcp{} cuts, a large amount of impact parameter mixing occurs
leading ultimately to a finite limit on the true centrality of any
selected sample of events. We can now quantify this saturation by
providing an estimate of $\langle b\rangle$ for any selection of
data, without relying on any specific model of collisions. We present
here the mean values of reduced impact parameter for cuts corresponding
to experimental centralities $c_{X}<10\%$ or $c_{X}<1\%$ performed
using the total transverse energy of LCP, \etlcp. It should be noted
that equivalent results are found using the \mtot{} observable.

\begin{figure}
\includegraphics[clip,width=1\columnwidth]{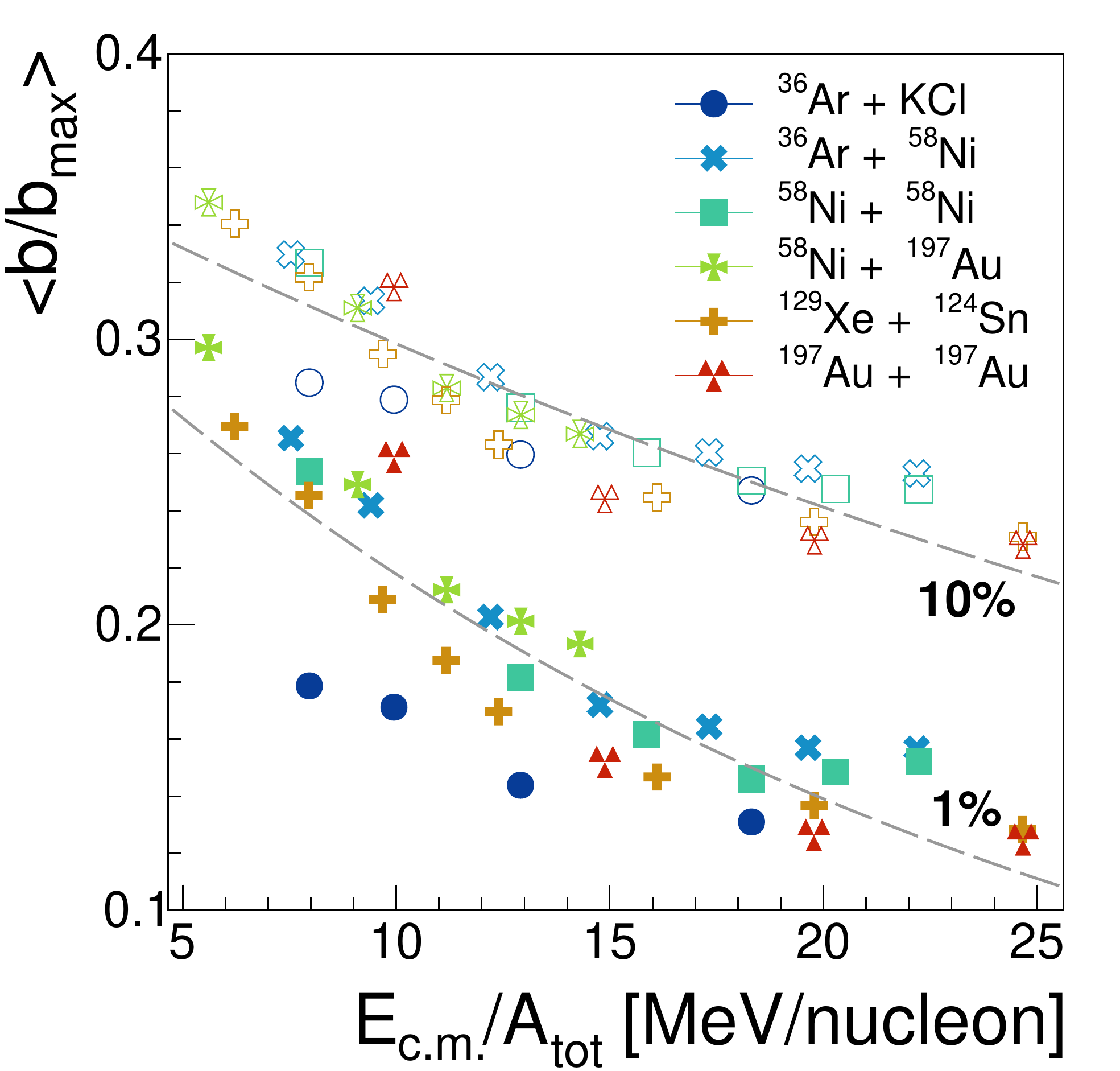}

\caption{Mean reduced impact parameter $\langle b/b_{\mathrm{max}}\rangle$
for central collisions selected with a 10\% (open symbols) or 1\%
(full symbols) centrality cut using \etlcp, as a function of available
center of mass energy per nucleon, $E_{\mathrm{c.m.}}/A_{\mathrm{tot}}$.
Dashed curves are to guide the eye.\label{fig:how-central-are-they}}
\end{figure}

Figure \ref{fig:how-central-are-they} shows the mean value of $b/b_{\mathrm{max}}$
for the impact parameter distributions we have reconstructed for all
systems and bombarding energies and for the two cuts, as a function
of the available center of mass energy of each reaction. The values
of $\langle b/b_{\mathrm{max}}\rangle$ can be seen to cluster around
a near-universal energy dependency for each cut, which appears to
be independent of system mass and entrance channel asymmetry. In each
case, $\langle b/b_{\mathrm{max}}\rangle$ decreases with increasing
center of mass energy. With the more restrictive cut of $c_{X}<1\%$,
the selectivity of the cuts can be seen to increase significantly,
especially for the highest energies. Values of $0.35\geq\langle b/b_{\mathrm{max}}\rangle\geq0.24$
are obtained for the 10\% centrality cut, and $0.29\geq\langle b/b_{\mathrm{max}}\rangle\geq0.14$
with the 1\% centrality cut.

It is instructive to compare these results with the predictions of
the sharp cut-off approximation (SCA) for the same cuts (see the end
of Sec. \ref{subsec:Application-to-simulated}). The SCA would expect
an energy-independent upper limit of $b_{\mathrm{cut}}/b_{\mathrm{max}}\approx0.3$
for the 10\% cut, or $b_{\mathrm{cut}}/b_{\mathrm{max}}=0.1$ with
a 1\% centrality cut. For a geometric (triangular) impact parameter
distribution, the mean value is simply related to the cut-off by $\langle b^{\mathrm{SCA}}\rangle=\nicefrac{2}{3}b_{\mathrm{cut}}$,
therefore the mean values in Figure \ref{fig:how-central-are-they}
should be compared to $\langle b^{SCA}/b_{\mathrm{max}}\rangle\approx0.1$
in the case of the 10\% centrality cut, and $\langle b^{SCA}/b_{\mathrm{max}}\rangle\approx0.07$
for $c_{X}<1\%$. It is clear that impact parameter estimation using
the SCA for central collisions greatly underestimates representative
values of $\langle b\rangle$ for the selected data samples. The method
presented in this paper can characterize data not only in terms of
far more representative mean values of $b$, but indeed provide an
estimation of the actual impact parameter distribution which could
be used as input to a transport model calculation.

\begin{table}
\begin{tabular}{ccr@{\extracolsep{0pt}.}lr@{\extracolsep{0pt}.}lr@{\extracolsep{0pt}.}lr@{\extracolsep{0pt}.}lr@{\extracolsep{0pt}.}lr@{\extracolsep{0pt}.}l}
System & $E_{\mathrm{proj}}$ & \multicolumn{2}{c}{$\alpha$} & \multicolumn{2}{c}{$\gamma$} & \multicolumn{2}{c}{$\theta$} & \multicolumn{2}{c}{$X_{\mathrm{max}}$} & \multicolumn{2}{c}{$X_{\mathrm{min}}$} & \multicolumn{2}{c}{$\chi^{2}$}\tabularnewline
 & {[}MeV/nucleon{]} & \multicolumn{2}{c}{} & \multicolumn{2}{c}{} & \multicolumn{2}{c}{} & \multicolumn{2}{c}{} & \multicolumn{2}{c}{} & \multicolumn{2}{c}{}\tabularnewline
\hline 
\multirow{4}{*}{\col{Ar}{36}{KCl}{}} & 32 & 0&95 & 1&25 & 0&20 & 14&1 & 3&7 & 13&2\tabularnewline
 & 40 & 0&98 & 1&46 & 0&23 & 15&6 & 4&2 & 6&5\tabularnewline
 & 52 & 0&88 & 1&40 & 0&22 & 17&6 & 4&0 & 8&2\tabularnewline
 & 74 & 0&89 & 1&60 & 0&21 & 19&8 & 4&3 & 10&2\tabularnewline
\hline 
\multirow{7}{*}{\col{Ar}{36}{Ni}{58}} & 32 & 1&37 & 1&12 & 0&25 & 14&8 & 2&5 & 85&4\tabularnewline
 & 40 & 1&23 & 1&14 & 0&27 & 16&8 & 2&6 & 46&4\tabularnewline
 & 52 & 1&07 & 1&17 & 0&28 & 19&1 & 2&7 & 53&5\tabularnewline
 & 63 & 0&99 & 1&19 & 0&29 & 20&6 & 2&7 & 28&1\tabularnewline
 & 74 & 0&95 & 1&21 & 0&27 & 22&4 & 3&8 & 43&8\tabularnewline
 & 84 & 0&96 & 1&24 & 0&27 & 23&2 & 3&8 & 25&8\tabularnewline
 & 95 & 0&98 & 1&28 & 0&27 & 24&1 & 3&9 & 56&3\tabularnewline
\hline 
\multirow{6}{*}{\col{Ni}{58}{Ni}{58}} & 32 & 1&24 & 1&19 & 0&28 & 16&3 & 3&7 & 55&4\tabularnewline
 & 52 & 0&97 & 1&15 & 0&30 & 21&6 & 3&8 & 14&0\tabularnewline
 & 63 & 0&92 & 1&16 & 0&28 & 24&1 & 3&8 & 5&9\tabularnewline
 & 74 & 0&81 & 1&17 & 0&27 & 25&8 & 3&7 & 4&2\tabularnewline
 & 82 & 0&96 & 1&45 & 0&29 & 26&4 & 4&6 & 8&1\tabularnewline
 & 90 & 0&93 & 1&40 & 0&28 & 27&4 & 4&2 & 38&2\tabularnewline
\hline 
\multirow{5}{*}{\col{Ni}{58}{Au}{197}} & 32 & 1&79 & 1&79 & 0&34 & 16&2 & 4&1 & 84&3\tabularnewline
 & 52 & 1&63 & 1&66 & 0&37 & 23&0 & 4&2 & 49&5\tabularnewline
 & 64 & 1&48 & 1&64 & 0&37 & 26&4 & 4&3 & 15&1\tabularnewline
 & 74 & 1&44 & 1&62 & 0&39 & 28&4 & 4&1 & 45&6\tabularnewline
 & 82 & 1&51 & 1&85 & 0&41 & 29&7 & 4&7 & 43&2\tabularnewline
\hline 
\multirow{5}{*}{\xesn} & 25 & 1&26 & 0&95 & 0&32 & 19&6 & 2&8 & 63&5\tabularnewline
 & 32 & 1&24 & 1&08 & 0&34 & 24&0 & 3&2 & 40&3\tabularnewline
 & 39 & 1&18 & 1&17 & 0&34 & 28&4 & 3&5 & 55&1\tabularnewline
 & 45 & 1&14 & 1&23 & 0&34 & 31&2 & 3&8 & 56&5\tabularnewline
 & 50 & 1&14 & 1&35 & 0&34 & 34&1 & 4&0 & 36&6\tabularnewline
\hline 
\multirow{3}{*}{\xesngsi} & 65 & 1&09 & 1&40 & 0&36 & 38&4 & 2&9 & 2&0\tabularnewline
 & 80 & 1&11 & 1&50 & 0&34 & 42&5 & 3&1 & 1&5\tabularnewline
 & 100 & 1&18 & 1&64 & 0&38 & 45&2 & 3&5 & 2&7\tabularnewline
\hline 
\multirow{4}{*}{\col{Au}{197}{Au}{197}} & 40 & 1&23 & 1&27 & 0&42 & 35&5 & 2&1 & 30&1\tabularnewline
 & 60 & 1&22 & 1&62 & 0&45 & 47&8 & 1&2 & 68&1\tabularnewline
 & 80 & 1&24 & 1&61 & 0&45 & 54&9 & 2&8 & 16&9\tabularnewline
 & 100 & 1&26 & 1&64 & 0&49 & 58&5 & 3&1 & 52&0\tabularnewline
\end{tabular}

\caption{Parameters of fits to total charged particle multiplicity distributions
$P(N_{C})$ for all datasets. See Sec. \ref{subsec:Specific-implementation}
for meaning of parameters. $\chi^{2}$ is the reduced chi-square value
of each fit.\label{tab:multiplicity-fits}}
\end{table}
\begin{table}
\begin{tabular}{ccr@{\extracolsep{0pt}.}lr@{\extracolsep{0pt}.}lr@{\extracolsep{0pt}.}lrrr@{\extracolsep{0pt}.}l}
System & $E_{\mathrm{proj}}$ & \multicolumn{2}{c}{$\alpha$} & \multicolumn{2}{c}{$\gamma$} & \multicolumn{2}{c}{$\theta$} & $X_{\mathrm{max}}$ & $X_{\mathrm{min}}$ & \multicolumn{2}{c}{$\chi^{2}$}\tabularnewline
 & {[}MeV/nucleon{]} & \multicolumn{2}{c}{} & \multicolumn{2}{c}{} & \multicolumn{2}{c}{{[}MeV{]}} & {[}MeV{]} & {[}MeV{]} & \multicolumn{2}{c}{}\tabularnewline
\hline 
\multirow{4}{*}{\col{Ar}{36}{KCl}{}} & 32 & 0&35 & 0&76 & 6&1 & 162 & 3 & 1&2\tabularnewline
 & 40 & 0&37 & 0&89 & 7&5 & 196 & 8 & 1&0\tabularnewline
 & 52 & 0&35 & 1&02 & 8&5 & 269 & 12 & 1&2\tabularnewline
 & 74 & 0&40 & 1&32 & 11&8 & 389 & 19 & 3&7\tabularnewline
\hline 
\multirow{7}{*}{\col{Ar}{36}{Ni}{58}} & 32 & 0&97 & 1&17 & 8&5 & 148 & 9 & 3&3\tabularnewline
 & 40 & 0&83 & 1&17 & 10&0 & 183 & 11 & 2&5\tabularnewline
 & 52 & 0&68 & 1&26 & 12&0 & 251 & 15 & 2&3\tabularnewline
 & 63 & 0&60 & 1&35 & 13&1 & 328 & 18 & 1&3\tabularnewline
 & 74 & 0&60 & 1&46 & 14&9 & 402 & 30 & 1&8\tabularnewline
 & 84 & 0&60 & 1&52 & 16&1 & 463 & 32 & 1&7\tabularnewline
 & 95 & 0&62 & 1&63 & 18&6 & 528 & 35 & 2&1\tabularnewline
\hline 
\multirow{6}{*}{\col{Ni}{58}{Ni}{58}} & 32 & 0&79 & 1&04 & 9&9 & 186 & 21 & 1&6\tabularnewline
 & 52 & 0&56 & 1&15 & 13&0 & 340 & 29 & 2&3\tabularnewline
 & 64 & 0&55 & 1&30 & 14&9 & 443 & 33 & 1&7\tabularnewline
 & 74 & 0&52 & 1&40 & 16&7 & 541 & 37 & 2&8\tabularnewline
 & 82 & 0&61 & 1&68 & 18&5 & 591 & 46 & 1&7\tabularnewline
 & 90 & 0&68 & 1&93 & 20&9 & 629 & 56 & 1&4\tabularnewline
\hline 
\multirow{5}{*}{\col{Ni}{58}{Au}{197}} & 32 & 1&41 & 1&71 & 12&8 & 223 & 32 & 6&7\tabularnewline
 & 52 & 1&08 & 1&45 & 17&2 & 391 & 36 & 6&1\tabularnewline
 & 64 & 0&93 & 1&50 & 18&8 & 521 & 39 & 3&8\tabularnewline
 & 74 & 0&92 & 1&63 & 21&6 & 630 & 44 & 4&8\tabularnewline
 & 82 & 0&92 & 1&74 & 23&3 & 716 & 48 & 5&8\tabularnewline
\hline 
\multirow{5}{*}{\xesn} & 25 & 0&74 & 0&68 & 11&1 & 241 & 6 & 2&2\tabularnewline
 & 32 & 0&67 & 0&69 & 12&4 & 310 & 5 & 1&7\tabularnewline
 & 39 & 0&57 & 0&75 & 13&9 & 408 & 7 & 2&9\tabularnewline
 & 45 & 0&55 & 0&89 & 15&4 & 496 & 24 & 1&6\tabularnewline
 & 50 & 0&57 & 1&06 & 16&0 & 584 & 34 & 1&3\tabularnewline
\hline 
\multirow{3}{*}{\xesngsi} & 65 & 0&59 & 1&32 & 19&9 & 822 & 34 & 1&5\tabularnewline
 & 80 & 0&61 & 1&52 & 23&5 & 1071 & 41 & 2&0\tabularnewline
 & 100 & 0&62 & 1&65 & 26&8 & 1374 & 44 & 3&6\tabularnewline
\hline 
\multirow{4}{*}{\col{Au}{197}{Au}{197}} & 40 & 1&07 & 1&23 & 24&0 & 521 & 18 & 24&4\tabularnewline
 & 60 & 0&67 & 1&38 & 25&5 & 1089 & 5 & 24&6\tabularnewline
 & 80 & 0&62 & 1&47 & 26&6 & 1648 & 31 & 3&0\tabularnewline
 & 100 & 0&65 & 1&68 & 34&6 & 2054 & 55 & 4&4\tabularnewline
\end{tabular}

\caption{Results of fits to total transverse LCP energy distributions $P(E_{t12})$
for all datasets. See Sec. \ref{subsec:Specific-implementation} for
meaning of parameters. $\chi^{2}$ is the reduced chi-square value
of each fit. \label{tab:et12-fits}}
\end{table}

\section{Conclusions}

One way to improve constraints on the nuclear equation of state from
comparisons between data on intermediate energy heavy-ion collisions
and transport model calculations is by providing a model-independent
estimation of the impact parameter distribution representative of
any selected set of experimental data. To do so requires to explicitly
take into account the fluctuations in the relationship between any
observable $X$ and the impact parameter $b$, as first shown in \citep{Das2018Relating,Rogly2018Reconstructing}.
In this article we have shown how the method, first developed for
ultra-relativistic collisions, can be adapted and used in the \amev{20\textendash 100}
bombarding energy range. Notably, we have proposed a new parametrization
of the relationship between the mean value of an observable and the
impact parameter whose parameters are simple to interpret in terms
of the shape of this relationship.

We have shown, using a complete simulation of \col{Ni}{58}{Ni}{58}
collisions at \amev{52} measured by the INDRA array, calculated using
the AMD transport model coupled with the GEMINI++ statistical decay
code, that the method is capable of reconstructing the impact parameter
distributions associated to experimental events taking account of
secondary decay and finite detector acceptance effects. In this example,
the impact parameter distributions for a set of experimental cuts
defined over the full range of centralities were correctly reproduced,
and even for the most central cut for which the method shows a slight
underestimation of the actual impact parameters, the estimated impact
parameter distribution is far closer to the truth than an estimation
based on the commonly-used sharp cut-off approximation. These calculations
also indicate that one of the major assumptions of the method, that
the reduced variance of the observable is independent of centrality,
is a reasonable approximation, at least for the two observables we
have studied in this paper.

We then applied the approach to a very wide range of data for different
collisions measured with INDRA between \amev{25} and \amev{100},
where in each case the data were recorded according to a ``minimum-bias''
trigger based on a minimum number of fired telescopes over nearly
the full $4\pi$ solid angle around the target. Two commonly-used
observables have been employed, the total multiplicity of charged
products, \mtot, and the total transverse energy of light charged
particles with $Z\leq2$, \etlcp. Excellent fits to the inclusive
distributions $P(X)$ of each observable have been achieved for all
collisions, even if for \mtot{} the failure to reproduce the distributions
close to the minimum bias trigger (most peripheral collisions) biases
the apparent goodness-of-fit as measured by reduced $\chi^{2}$ values.

The parameters determined by the fits allow to deduce the joint probability
distribution $P(X,c_{b})$ from which impact parameter distributions
for any selection of data can be reconstructed, or distributions of
the observable for a given range of $b$. The relative fluctuation
of the joint probability distributions about the mean value of the
observable for each $b$ has been shown to differ according to the
observable, with the total multiplicity of charged products, \mtot{},
associated with sub-Poissonian fluctuations ($\theta<1$), whereas
the total transverse energy of light charged particles, \etlcp{},
exhibits much larger fluctuations ($\theta\gg1$), for all data studied
in this article.

The asymptotic values of the mean values of both observables for $b=0$
collisions can be extrapolated from the fit results, and may provide
new constraints for transport model calculations. We have shown, in
particular, that the total transverse energy of light charged particles
has mean values for head-on collisions which show a near-universal
dependence on the available energy in the center of mass of the collisions.
This result should be confronted with different microscopic model
predictions.

Finally, we have characterized the true centrality of a commonly-used
event selection employing high-\etlcp{} cuts to retain the ``most
central collisions''. The results are largely independent of total
system mass and mass-asymmetry of the entrance channel, each system
showing very similar evolution of the mean reduced impact parameter
$\langle b/b_{max}\rangle$ as a function of available energy for
each centrality cut. The actual representative mean values of reduced
impact parameters for these selections were shown to decrease with
increasing bombarding energy from 0.35 to 0.24 (for a 10\% centrality
cut), or from 0.29 to 0.14 (for a 1\% centrality cut), when the usual
sharp cut-off approximation (SCA) gives mean values of 0.2 or 0.07,
respectively, for these two centrality cuts, independently of the
reaction bombarding energy. This overestimation of the centrality
of each data sample would skew comparison with any transport model
by using the wrong impact parameters as input. This is why we have
tried to demonstrate in this paper that a new, model-independent method
for estimating the impact parameter distributions of selected experimental
events is feasible and should be used whenever possible in order to
improve the constraints that can be brought on the description of
nuclear dynamics and the nuclear equation of state by comparisons
between experimental data and microscopic transport model calculations.
\begin{acknowledgments}
We would like to thank all the technical staff of GANIL for their
continued support in performing the experiments. We gratefully acknowledge
support from the CNRS/IN2P3 Computing Center (Lyon - France) for providing
computing and data-processing resources needed for this work. We would
also like to thank the ROOT \citep{Brun1997ROOT} development team
without whose software none of the analyses would be possible. We
acknowledge support from R\'{e}gion Normandie under the R\'{e}seau d'Int\'{e}r\^{e}t
Normand FIDNEOS (RIN/FIDNEOS). We would also like to signal that the
software necessary to perform the analyses presented in this paper
will be made available to the entire community as part of the KaliVeda
heavy-ion analysis toolkit \citep{KaliVeda} upon publication. 
\end{acknowledgments}

\appendix

\section{Inclusive impact parameter distributions for INDRA data\label{sec:Inclusive-impact-parameter}}

In order to transform deduced centrality distributions $P(c_{b}|S)$
into impact parameter distributions using Eq. \ref{eq:general-cent-dist-to-ip-dist-transform}
requires to calculate the centrality for each impact parameter, $c_{b}(b)$,
(Eq. \ref{eq:realistic-pinel-centrality}) and deduce the value of
$b_{0}$ from the (measured) total reaction cross-section by numerical
inversion of Eq. \ref{eq:realistic_pinel_sigma}, assuming a typical
value of $\Delta b\approx0.4$ fm.

\subsection{Analytic expression for total cross-section \label{subsec:Analytic-expression-for-xsec}}

To calculate the total reaction cross-section for a given set of parameters
$b_{0}$ and $\Delta b$, we have, by definition,
\[
\sigma_{R}=\int_{0}^{\infty}2\pi b\left[1+\exp\left(\frac{b-b_{0}}{\Delta b}\right)\right]^{-1}\,\mathrm{d}b
\]
and making the substitutions $b=t\Delta b$ and $b_{0}=x\Delta b$
we arrive at
\[
\sigma_{R}=2\pi(\Delta b)^{2}\int_{0}^{\infty}\frac{t}{1+\exp\left(t-x\right)}\,\mathrm{d}t
\]
This definite integral is related to the complete Fermi-Dirac integral
\begin{equation}
F_{j}(x)=\frac{1}{\Gamma(j+1)}\int_{0}^{\infty}\frac{t^{j}}{1+\exp\left(t-x\right)}\,\mathrm{d}t\label{eq:complete-FD-integral}
\end{equation}
with $j=1$, where $\Gamma(j+1)$ is the gamma function, $\Gamma(j+1)=j!$
for integer $j$. In general the value of this integral is given by
a polylogarithm, $\mathrm{Li}_{s}(z)$, specifically
\[
F_{j}(x)=-\mathrm{Li}{}_{j+1}(-\mathrm{e}^{x})
\]
and in this particular case by the negative dilogarithm, $-\mathrm{Li}_{2}(-e^{x})$.
Therefore we have for the final expression of the total cross-section
which normalizes correctly the probability distribution of Eq. \eqref{eq:realistic-b-dist},
\begin{equation}
\sigma_{R}=-2\pi(\Delta b)^{2}\mathrm{Li}_{2}\left(-\exp\left(\frac{b_{0}}{\Delta b}\right)\right)\label{eq:realistic_pinel_sigma}
\end{equation}
This expression can be used to find $b_{0}$ for a given total cross-section
and width parameter $\Delta b$, by numerical inversion \citep{Brun1997ROOT}.

\subsection{Analytic expression for centrality \label{subsec:Analytic-expression-for-centrality}}

To calculate the centrality $c_{b}(b)$ we substitute Eq. \eqref{eq:realistic-b-dist}
into Eq. \eqref{eq:b-centrality}, and making the same substitutions
as above ($b=t\Delta b$, $b_{0}=x\Delta b$ ) we find
\[
c_{b}(b)=\frac{2\pi(\Delta b)^{2}}{\sigma_{R}}\int_{0}^{b/\Delta b}\frac{t'}{1+\exp\left(t'-x\right)}\,\mathrm{d}t'
\]
This definite integral can be calculated using the incomplete Fermi-Dirac
integral 
\[
F_{j}(a,x)=\frac{1}{\Gamma\left(j+1\right)}\int_{a}^{\infty}\frac{t^{j}}{1+\exp\left(t-x\right)}\,\mathrm{d}t,\;a\geq0
\]
with $a=b/\Delta b$, and the complete Fermi-Dirac integral $F_{j}(x)$
of Eq. \eqref{eq:complete-FD-integral}:
\[
\int_{0}^{a}\frac{t^{j}}{1+\exp\left(t-x\right)}\,\mathrm{d}t=\Gamma\left(j+1\right)\left[F_{j}\left(x\right)-F_{j}\left(a,x\right)\right]
\]
With $j=1$, $F_{1}(x)=-\mathrm{Li_{2}}(-e^{x})$ as above, while
the incomplete FD integral can be written (by integration by parts)
as

\[
F_{1}(a,x)=\frac{\pi^{2}}{6}-\frac{(a^{2}-x^{2})}{2}+a\ln\left(1+\mathrm{e}^{(a-x)}\right)+\mathrm{Li}_{2}\left(-\mathrm{e}^{(a-x)}\right)
\]
The final expression for the centrality is therefore

\begin{eqnarray}
c_{b}(b) & = & \frac{2\pi(\Delta b)^{2}}{\sigma_{R}}\left[\mathrm{-Li}_{2}\left(-\exp\left(\frac{b_{0}}{\Delta b}\right)\right)-\frac{\pi^{2}}{6}+\frac{(b^{2}-b_{0}^{2})}{2(\Delta b)^{2}}\right.\label{eq:realistic-pinel-centrality}\\
 &  & \left.-\frac{b}{\Delta b}\ln\left(1+\exp\left((b-b_{0})/\Delta b\right)\right)-\mathrm{Li}_{2}\left(-\mathrm{e}^{(b-b_{0})/\Delta b}\right)\right]
\end{eqnarray}

\end{document}